\documentclass[a4paper,10pt,aps,amsfonts,amsmath,nofootinbib,showpacs,byrevtex,showkeys%
,notitlepage
,reprint
]{revtex4-1}

\usepackage{hyperref,graphicx,bm}
\usepackage{subfigure}
\usepackage{bbm}

\newcommand{\cN}{{\cal N}}
\newcommand{\cD}{{\cal D}}
\newcommand{\cM}{{\cal M}}
\newcommand{\cP}{{\cal P}}
\newcommand{\cO}{{\cal O}}
\newcommand{\cA}{{\cal A}}
\newcommand{\cL}{{\cal L}}

\newcommand{\be}{\begin{equation}}
\newcommand{\ee}{\end{equation}}

\newcommand{\bea}{\begin{eqnarray}}
\newcommand{\eea}{\end{eqnarray}}

\providecommand*{\coloneq}{\mathrel{\mathop:}=}
\providecommand*{\eqcolon}{=\mathrel{\mathop:}}

\newcommand{\rk}{\right)}
\newcommand{\lk}{\left(}

\DeclareMathOperator{\Det}{Det}
\DeclareMathOperator{\Tr}{Tr}
\DeclareMathOperator{\tr}{tr}

\newcommand{\bra}[1]{\langle #1\rvert}
\newcommand{\ket}[1]{\lvert#1\rangle}

\newcommand*{\vev}[1]{\left< #1 \right>}
\def\abs#1{\lvert#1\rvert}

\renewcommand{\i}{\mathrm{i}}
\newcommand{\e}{\mathrm{e}}
\newcommand{\Id}{\mathbbm{1}}
\newcommand{\mR}{\mathbbm{R}}

\renewcommand*{\vec}[1]{\bm{\mathrm{#1}}}

\newcommand{\vC}{\vec{C}}
\newcommand{\vB}{\vec{B}}
\newcommand{\ve}{\vec{e}}
\newcommand{\vA}{\vec{A}}
\newcommand{\vx}{\vec{x}}
\newcommand{\vy}{\vec{y}}
\newcommand{\vD}{\vec{D}}

\newcommand{\vp}{\vec{p}}
\newcommand{\va}{\vec{a}}
\newcommand{\vj}{\vec{j}}
\newcommand{\vq}{\vec{q}}
\newcommand{\vd}{\vec{d}}
\newcommand{\vsigma}{\vec{\sigma}}
\newcommand{\vmu}{\vec{\mu}}

\renewcommand*{\d}[1][]{\mathop{\mathrm{d}^{#1}}\mkern-4mu}
\newcommand*{\dbar}[1][]{\mathop{\mathrm{d}\mkern-7mu\mathchar'26\mkern-1mu^{#1}}\mkern-4mu}

\begin{document}

\title{The effective potential of the confinement order parameter in the Hamiltonian approach}

\author{Hugo Reinhardt}
\author{Jan Heffner}
\affiliation{Institut f\"ur Theoretische Physik, Universit\"at T\"ubingen,
Auf der Morgenstelle 14, 72076 T\"ubingen, Germany}
\date{\today}

\begin{abstract}
The effective potential of the order parameter for confinement is calculated for SU($N$) Yang--Mills theory in the Hamiltonian approach. 
Compactifying one spatial dimension and using a background gauge fixing, this potential
is obtained within a variational approach by minimizing the energy density for given background field. In this formulation the
inverse length of the compactified dimension represents the temperature. Using Gaussian trial 
wave functionals we establish an analytic relation between the propagators in the background gauge at 
finite temperature and the corresponding zero-temperature propagators in Coulomb gauge. In the simplest truncation, neglecting the ghost
and using the ultraviolet form of the gluon energy, we recover the Weiss potential. Neglecting the ghost and using for the gluon energy $\omega (p)$ the approximate Gribov formula $\omega (p) \simeq p + M^2/p$ one finds a critical temperature 
of $\sqrt{3} M/\pi$. We explicitly show that the omission of the ghost drastically increases the transition temperature. From the full
non-perturbative
potential (with the ghost included) we extract a critical temperature of the deconfinement phase transition of $269$ MeV for the
gauge group SU($2$) and $283$ MeV for SU($3$).
\end{abstract}

\maketitle
\section{Introduction}

One of the major challenges of particle physics is the understanding of the deconfinement phase transition from the 
confined hadronic phase with chiral symmetry spontaneously broken to the deconfined quark gluon plasma with chiral symmetry
restored. This transition is expected to be driven by the gluon dynamics, which suggests to investigate this transition
first in pure Yang--Mills theory. In the absence of quarks the deconfinement phase transition is related to the center
of the gauge group \cite{Svetitsky:1982gs}: Center symmetry is realized in the low-temperature confining phase and spontaneously broken in the
high-temperature deconfining phase. When quarks are included, center symmetry is explicitly broken and the deconfinement phase transition
is expected to become a crossover. 

Understanding the deconfinement phase transition requires non-perturbative methods. In quenched QCD reliable results have been 
obtained by means of the lattice simulations \cite{Karsch:2001cy}. These methods become, however, extremely expensive when dynamical quarks are included and 
fail at large baryon densities due to the notorious fermion sign problem at non-vanishing chemical potential. Alternative 
non-perturbative methods, which are based on the continuum formulation of QCD are therefore desirable. In recent years substantial
progress has been made within continuum approaches to QCD \cite{Fischer:2006ub,Pawlowski:2005xe,Watson:2006yq,Feuchter:2004mk}.
 Among these is a variational approach to the Hamiltonian formulation of
Yang--Mills theory in Coulomb gauge \cite{Schutte:1985sd,Szczepaniak:2001rg,Feuchter:2004mk}. 
In this approach the energy density is minimized using Gaussian type ans\"atze for the Yang--Mills
vacuum wave functional. Within the approach of Ref.~\cite{Feuchter:2004mk} a decent description of the infrared sector of Yang--Mills theory 
was obtained \cite{Epple:2006hv,Schleifenbaum:2006bq,Reinhardt:2007wh,R11,Campagnari:2010wc,Campagnari:2011bk}.
Extension of this approach to 
full QCD has shown that the coupling of the quarks to the transversal gluons amplifies the spontaneous
breaking of chiral symmetry \cite{Pak:2011wu}. 
Recently this approach was also extended to finite temperatures by considering the grand canonical
ensemble, making a suitable quasi-particle ansatz for the density operator and minimizing the free energy \cite{Reinhardt:2011hq,R16}. 
In the present paper we present an alternative Hamiltonian approach 
to finite temperature Yang--Mills theory, which does not require an ansatz for the density operator. The finite temperature is introduced
here by compactifying one spatial dimension. From the physical point of view this approach therefore addresses, strictly speaking,
the Casimir problem. A summary of preliminary results obtained in this approach was reported in \cite{RX1}.

The ground state properties of a quantum field theory are usually studied by means of the Euclidean space
functional integral. This approach, which is also the basis for lattice calculations, can be easily extended to
finite temperatures by compactifying the Euclidean time axis, 
whence the momentum variable (energy)
corresponding to the compactified Euclidean time becomes discrete with values
given
by multiples of $2 \pi$ times the temperature (Matsubara frequencies). Due to the O$(4)$ symmetry of Euclidean quantum field theory 
(at zero temperature) all four Euclidean dimensions are equivalent and instead of compactifying the 
time one can equally well compactify one of the spatial dimensions. This procedure can be also applied to the
Hamiltonian formulation of a quantum field theory. Thus, instead of considering the canonical or grand canonical 
ensemble of a theory, its extension
to finite temperatures can be equally well obtained by considering the theory for one compactified spatial 
dimension and identifying the length of the compactified dimension with the inverse temperature. 
In the present
paper we apply this procedure to
study Yang--Mills theory at finite temperature and, in particular, will investigate the
deconfinement phase transition. To make contact with the
previous Hamiltonian formulation of Yang--Mills theory in Coulomb gauge we use here a background gauge fixing, which
reduces to Coulomb gauge for vanishing background field. We will consider 
a constant Abelian background field living in the Cartan subalgebra and being directed along
the compact dimension. We will then calculate
the energy density as a function of the background field and the length of the compactified dimension. 
As will be shown in Sect.~\ref{sectionII-neu}
this quantity provides the finite-temperature effective potential of the Polyakov loop,
which is the order parameter of confinement. 
From this potential the
critical temperature of the deconfinement phase transition will be extracted. 

The organization of the paper is as follows:
In the next section we show how the effective potential of the Polyakov loop can be calculated in the Hamiltonian approach. In Sect.~\ref{sectionIIA} we discuss how a background field chose in the Cartan subalgebra and directed along a compactified dimension can figure as an order parameter for confinement alternativ to the Polyakov loop. In Sect.~\ref{sec3} we show how in the Hamiltonian approach a Lorentz invariant quantum field theory can be extanded to finite temperature by compactifying a spatial dimension. The upshot of Sect.~\ref{sectionII-neu} is that the effective potential of the confinement order parameter can be obtained in the Hamiltonian approach form the energy density calculated as a function of a constant background field living in the Cartan subalgebra and beeing directed along a compactified spatial dimension. The background field method is formulated in the Hamiltonian approach in Sect.~\ref{sectionIVback}. 
In Sect.~\ref{III} we develop the Hamiltonian approach in the presence of a background field in a background gauge. The 
crucial point is here the resolution of Gauss' law. In the absence of the background field this approach reduces to the
Hamiltonian approach in Coulomb gauge \cite{Feuchter:2004mk}. We also discuss here the choice of the variational wave functional 
required to ensure that the expectation
value of the gauge field agrees with the background field.
The Dyson--Schwinger equations (DSE) for the various propagators in the presence of a background field are derived in Sect.~\ref{IV}. In Sect.~\ref{V} we consider a constant background field living in the Cartan subalgebra of the gauge group and relate the solutions
of the corresponding DSEs to the ones of the Hamiltonian approach in Coulomb gauge in the 
absence of the background field.  For pedagogical reasons we develop the formalism for the gauge group SU($2$) and 
present its extension to SU($N$) in Sect.~\ref{sectVI-neu}. In Sect.~\ref{VG} we summarize the 
results obtained in the  Hamiltonian approach in Coulomb gauge for the various propagators, which serve as input
for the background field calculation of the effective potential of the Polyakov loop given in Sect.~\ref{sectVI}. A short summary and our conclusions are given in Sect.~\ref{VII}. Some lengthy mathematical derivations are deferred to appendices.

\section{\label{sectionII-neu}The effective potential of the confinement order parameter in the Hamiltonian approach}

\subsection{\label{sectionIIA} The Polyakov loop potential}

Consider Yang--Mills theory at finite temperatures and assume that the temperature is introduced by 
compactifying the euclidean time axis, identifying the length $L$ of the compactified dimension
 with the inverse temperature. In the absence of fermions the finite-temperature Yang--Mills
theory is invariant under gauge transformations $U(x_0, \vx)$, which are temporally periodic up to an element $z_k$
of the center Z$(N)$ of the gauge group SU$(N)$
\be
\label{189-a1}
U (L, \vx) = z_k (\vx) U (0, \vx\ )  ,
\ee
where the center element is given by
\be
\label{209-a4}
z_k = \e^{-2 \pi \tilde{\mu}_k} = \e^{\i \frac{2 \pi}{N} k} \Id \in \text{Z}(N), \quad k = 0, 1, \dots, N - 1 
\ee
with $\tilde{\mu}_k = \tilde{\mu}^a_k T_a$ being a co-weight vector. (We use antihermitean generators $T_a$.)
The order parameter of confinement is 
the expectation value $\langle P  [A_0] (\vx) \rangle$ of 
the Polyakov loop \cite{Polyakov:1978vu}
\be
\label{194-a2}
P [A_0] (\vx) = \frac{1}{N} \tr \cP \exp \lk - \int^L_0 \d[]x_0 A_0 (x_0, \vx) \rk  \, .
\ee
This quantity can be related to  the free energy $F_q (\vx)$ 
of an isolated static (infinitely heavy) quark at space position $\vx$ by \cite{Svetitsky:1982gs}\footnote{
The phases of Yang--Mills theory can be alternatively distinguised by the ``disorder'' parameter of confinement, the 't~Hooft loop \cite{R27}, which is dual to the temporal Wilson loop but can be used both at zero and finite temperature. Its continuum representation was derived in Ref.~\cite{R28}. Evaluation of the 't~Hooft loop in the Hamiltonian approach in Coulomb gauge (at zero temperature) has given a perimenter law \cite {Reinhardt:2007wh} as expected for the confined phase.}
\be
\label{244-3}
\langle P [A_0] ( \vx) \rangle = \e^{- L F_q (\vx)}
\ee
and
 is entirely defined within 
pure Yang--Mills theory. Under gauge transformations satisfying 
Eq.~(\ref{189-a1}) the Polyakov loop transforms as 
\be
\label{200-a3}
P [A_0] (\vx) \to z_k P [A_0] (\vx) \, 
\ee
and so does $\langle P  [A_0] \rangle \to z_k \langle P  [A_0] \rangle$. 
It follows that
the order parameter $\langle P [A_0] \rangle$ vanishes in the center symmetric confining phase, while $\langle P [A_0] \rangle 
\neq 0$ in the deconfining phase with center symmetry broken.

In the continuum theory the Polyakov loop is most easily calculated in Polyakov gauge, defined by 
$\partial_0 A_0 = 0$ and $A_0$ living in the Cartan subalgebra.\footnote{In Polyakov gauge
topological defects like magnetic monopoles and center vortices occur \cite{Reinhardt:1997rm, RX3}, which 
presumably are the ``confiner'' of the theory.} 
In this gauge the path ordering in Eq.~(\ref{194-a2}) becomes irrelevant
\be
\label{238-a8}
P [A_0] 
(\vx) = \frac{1}{N} \tr \e^{- L A_0(\vx)}
\ee
and under a global center transformation (\ref{200-a3}) $A_0$ transforms as 
\be
\label{249-a10}
A_0 \to A_0 + \frac{2 \pi}{L} \tilde{\mu}_k \, .
\ee
In Refs.~\cite{Marhauser:2008fz,Braun:2007bx} 
it was shown that in this gauge the inequality
\be
\label{2}
\langle P [A_0] \rangle \leq P [\langle A_0 \rangle ]
\ee
holds, and furthermore that also $P [\langle A_0 \rangle]$ vanishes in the 
confining phase. This establishes $P [\langle A_0 \rangle]$ (alternative to $\langle P [A_0] \rangle$) as an order parameter of confinement. 
Furthermore, due to the unique relation (\ref{238-a8}) between the gauge field $A_0$ in Polyakov gauge and $P[A_0]$, instead of $P [\langle A_0 \rangle]$ one can also use $\langle A_0 \rangle$ itself as alternative order parameter of confinement. As a consequence the phase structure of finite temperature Yang--Mills theory can be extracted from the effective potential $e [\langle A_0 \rangle ]$ of $\langle A_0 \rangle$. By gauge invariance a non-vanishing $\langle A_0 \rangle$ requires the presence of an external background field $a$. Choosing
$a$ to be constant and diagonal (thus being in the Polyakov gauge) and to satisfy $a = \langle A_0 \rangle$ the 
background field $a$ becomes an order parameter of confinement, whose value is determined by the minimum of the 
effective potential $e [ \langle A_0 \rangle = a]$. Gauge invariance ensures that this potential is invariant under the Z$(N)$ transformation (\ref{249-a10})
\be
\label{305-6}
e \left[\langle A_0 \rangle + \frac{2 \pi}{L} \tilde{\mu}_k \right] = e \left[ \langle A_0 \rangle \right] \, .
\ee
This potential was first calculated in one loop perturbation theory in Ref.~\cite{Weiss:1980rj}. 
Corrections to this potential from the renormalization
of the kinetic terms of the gauge field were calculated in Ref.~\cite{Engelhardt:1997pi}. 
It was found that $e [a]$ has its minimum at $\bar{a} = 0$, so that $\langle P \rangle \simeq P [\bar{a}] = 1$, 
implying that the theory is in the deconfining
phase with center symmetry broken. This is the expected behavior at high temperature, where the one-loop calculation is reliable. 
In the low temperature confined phase one expects that $e [a]$ 
is minimal at ``center symmetric configurations'' $\bar{a}$, for which $\langle P \rangle \simeq P [\bar{a}] = 0$. 
Such a behavior was found
e.g. in Ref.~\cite{Braun:2007bx, RX2}, where the effective potential was
determined non-perturbatively from a renormalization group flow equation.

\subsection{\label{sec3}Finite temperature from compactification of a spatial dimension}

Obviously, the order parameter $\langle P \rangle \simeq P[\langle A_0 \rangle]$ or $\langle A_0 \rangle$ is not directly accessible
in Weyl gauge $A_0 = 0$, which is assumed in the canonical quantization. However, by O$(4)$
symmetry, all four Euclidean dimensions are equivalent and instead of compactifying the time, one can equally well 
introduce the temperature by compactifying one of the spatial dimension, say the $x_3$-axis, and consider $\langle A_3 \rangle$
as an order parameter for confinement. 
This can be seen as follows:

Consider Yang--Mills theory at finite temperature $L^{- 1}$, which is defined by the partition function
\be
\label{en-1}
Z (L) = \Tr \e^{- L H (\vA)} \, .
\ee
Here $H (\vA)$ is the usual Yang--Mills Hamiltonian defined by canonical quantization in Weyl gauge $A_0 = 0$.
The partition function (\ref{en-1}) can be equivalently represented by the Euclidean functional integral, see 
for example Ref.~\cite{Reinhardt:1997rm}
\be
\label{en2}
Z (L) = \int_{x^0 - \text{pbc}} \prod\limits_\mu \cD A_\mu (x) \, \e^{- S [A]} \, ,
\ee
where
\be
\label{en-3}
S [A] = \int^{L/2}_{- L/2} \d x^0 \int \d[3] x \, \cL \lk A^\mu ; x^\mu \rk
\ee
is the Euclidean action and the functional integration is performed over temporally periodic fields
\be
\label{en-190}
A^\mu \lk L/2, \vx \rk = A^\mu \lk - L/2, \vx \rk \, ,
\ee
 which is
indicated in Eq.~(\ref{en2}) by the subscript $x^0 - \text{pbc}$. This boundary condition is absolutely necessary at finite
$L$ but becomes irrelevant in the zero temperature $(L \to \infty)$ limit. 

We perform now the following change of variables 
\begin{align}
\label{en-5}
x^0 \to z^3 & \quad \quad A^0 \to C^3 \nonumber\\
x^1 \to z^0 & \quad \quad A^1 \to C^0 \nonumber\\
x^2 \to z^1 & \quad \quad A^2 \to C^1 \nonumber\\
x^3 \to z^2 & \quad \quad A^3 \to C^2 \, .
\end{align}
Due to the O$(4)$ invariance of the Euclidean Lagrangian we have 
\be
\label{en-6}
\cL \lk A^\mu, x^\mu \rk = \cL \lk C^\mu, z^\mu \rk
\ee
and the partition function (\ref{en2}) can be rewritten as 
\be
\label{en-7}
Z (L) = \int_{z^3 - \text{pbc}} \prod_\mu \cD C^\mu (z) \, \e^{- \tilde{S} [C^\mu]} \, ,
\ee
where the action is now given by
\be
\label{en-8}
\tilde{S} [C^\mu] = \int \d z^0 \d z^1 \d z^2 \int^{L/2}_{- L/2} \d z^3 \cL \lk C^\mu, z^\mu \rk 
\ee
and the functional integration runs over fields satisfying periodic boundary condition in the $z^3-$direction
\be
\label{en-9}
C^\mu \lk z^0, z^1, z^2, L/2 \rk = C^\mu \lk z^0, z^1, z^2, - L/2 \rk \, .
\ee
We can now interprete $z^0$ as time and $\vec{z} = \lk z^1, z^2, z^3 \rk$ as space coordinates and perform a usual canonical 
quantization in ``Weyl gauge'' $C^0 = 0$, interpreting $\vC = \lk C^1, C^2, C^3 \rk$ as spatial coordinates of the gauge field, which
are defined, however, not on $\mR^3$ but instead on $\mR^2 \times S^1$. We obtain then the usual Yang--Mills Hamiltonian in which, however, the integration over $z^3$ is restricted to the interval $\left[- \frac{L}{2}, \frac{L}{2} \right]$. Let us denote this Hamiltonian by $\tilde{H}(\vC,L)$. Obviously $\tilde{H}(\vC ,L \to \infty ) =H(\vC)$.
Reversing the steps which lead from (\ref{en-1}) to (\ref{en2}) and taking into account the irrelevance of the temporal 
boundary conditions in the functional integral
for an infinite time-interval we obtain from Eq.~(\ref{en-7}) the alternative representation of the partition function
\be
\label{en-232}
Z (L) = \Tr \e^{- \int \d z^0 \tilde{H}(\vC,L)} = \lim\limits_{T \to \infty} \Tr \e^{- T \tilde{H}(\vC,L)} \, .
\ee
Due to the infinite $z^0$-(time-)interval $T \to \infty$ only the lowest eigenvalue of $\tilde{H}(\vC,L)$ contributes to the partition 
function $Z (L)$. The calculation of $Z (L)$ is thus reduced to solving the Schr\"odinger equation $\tilde{H}(\vC,L)\psi (\vC)= E \psi (\vC)$ for the vacuum state on the space manifold $\mR^2 \times S^1 (L)$, where $S^1 (L)$ is a circle with circumference $L$. 

The upshot of the above consideration is that finite-temperature gauge theory can be described in the Hamiltonian approach
by compactifying a spatial dimension and solving the corresponding Schr\"odinger equation for the vacuum sector. As the above derivation shows
this equivalence holds for any O$(4)$ invariant quantum field theory.

The above consideration for the partition function can be extended to the finite-temperature effective potential 
$e [\langle A_0 \rangle]$. One finds
that $e [\langle A_0 \rangle = a]$ can be calculated in the Hamiltonian approach from $e [\langle A_3 \rangle = a]$ with the $z^3$-axis
compactified. Furthermore, as shown in Ref.~\cite{WeinbV2}, in the Hamilton approach the effective potential $e [\langle A_3 \rangle = a]$ is
given by the energy density in the state that minimizes $\langle H \rangle$ for given $\langle A_3 \rangle$. 
In this paper we calculate the effective potential $e \left[ \langle A_3 \rangle = a 
\right]$ in the Hamiltonian approach using the representation (\ref{en-232}) of the partition function. In the spirit of the discussion given in Sect.~\ref{sectionIIA} we will refer to $e[\vev{A_3}]$ as the effective potential of the Polyakov loop.

A few comments are in order: In the present paper we 
use the Hamiltonian approach merely as a mathematical tool to calculate the effective potential of the
background field at finite temperatures. As the previous considerations show, strictly speaking, time in the present approach 
is not the true  time but a spatial dimension
and the compactified spatial dimensions represents the time. Alternatively, interpreting the time of the present 
Hamiltonian approach as 
the true time the present approach actually describes Yang--Mills theory in a space with one compactified spatial dimension and is
thus reminiscent of the Casimir problem. 

\section{\label{sectionIVback}The background field method in the Hamiltonian approach}

Let $\psi [A] = \langle A | \psi \rangle$ be the gauge invariant wave functional of the Yang--Mills vacuum. The
generating functional of the Hamiltonian approach is then defined by
\be
\label{ee-1} e^{W [j]} = \langle \psi | e^{\int \vj \vA} | \psi \rangle \, .
\ee
Fixing the gauge by the standard Faddeev--Popov method we have explicitly
\be
\label{ee-2}
e^{W [j]} = \int  \cD A  \, \Det \cM [A] \delta \lk f^a [A] \rk \e^{- S [A] + \int \vj \vA } \, .
\ee
Here we have rewritten the wave functional as
\be
\label{ee-3}
\psi [A] = \exp \lk - \frac{1}{2} S [A] \rk \, 
\ee
and assumed that it is real.
Furthermore
\be
\label{ee-4}
f^a [A] = 0
\ee
is the gauge fixing constraint and 
\be
\label{ee-5}
\cM^{ab} = \frac{\delta f^a [A^U]}{\delta \Theta^b}
\ee
is the corresponding Faddeev--Popov kernel with 
\be
\label{ee-6}
A^U = U^\dagger (\partial + A) U = U^\dagger D U \,, \quad U = \exp (- \Theta)
\ee
being the gauge transformation.
Formally Eq.~(\ref{ee-2}) looks like the standard Yang--Mills functional integral of the Lagrangian formulation
in Euclidean space. However, in Eq.~(\ref{ee-2}) $S [A]$ is not the usual classical action and the functional
integral runs over the spatial components of the gauge field only. In principle, $S [A]$ is not known. However,
in recent years there has been substantial progress in  determining this wave functional approximately from a 
variational principle. Furthermore, in $2$+$1$ dimension more detailed information on the Yang--Mills vacuum wave 
functional has been obtained from lattice studies \cite{Greensite:2010yp} and strong coupling considerations \cite{Karabali:1998yq, Greensite:2011pj}. 
In the present paper
we will use for $S [A]$ the results of the variational calculations in Coulomb gauge carried out in Refs.~\cite{Feuchter:2004mk,Epple:2006hv,R16}. 
For the moment let us assume that $\psi [A]$ and thus $S [A]$ is known. The similarity of Eq.~(\ref{ee-2}) with
the standard Lagrangian Yang--Mills functional integral allows us to apply the standard background field method \cite{Pokorski:1987ed}
to simplify the evaluation of the effective action
\be
\label{ee-7}
\Gamma [\bar{A}] =  \left(   - W [j] + \int \vj \bar{\vA} \right)  _{j = j (\bar{A})} \, , \quad \bar{A} = \frac{\delta W [j]}{\delta j}\, .
\ee
For this purpose we split the gauge field $A$ into a background field $a$ and a fluctuating part $\cA$
\be
\label{ee-8}
A = a + \cA \, .
\ee
The gauge transformation (\ref{ee-6}) of the total gauge field $A$ can be distributed among the background field
$a$ and the fluctuation $\cA$ in different ways. The two common and convenient possibilities are:
\begin{enumerate}
\item quantum gauge transformations, which leave the background field invariant
\bea
\label{ee-9}
\delta_q a & = & 0 \nonumber\\
\delta_q \cA & =& [D, \delta \Theta] = [d + \cA, \delta \Theta] \, ,
\eea
where
\be
\label{ee-10}
d = \partial + a
\ee
is the covariant derivative with the background field
\item background gauge transformations
\bea
\label{ee-11}
\delta_b a & = & [d, \Theta] \nonumber\\
\delta_b \cA & = & [\cA, \Theta] \, .
\eea
\end{enumerate}
Following the standard background field method \cite{Pokorski:1987ed} we introduce the generating functional
\be
\label{ee-12}
\e ^ {\tilde{W} [a, j] } = \int \cD \cA \, \Det \tilde{\cM} [A] \delta \lk \tilde{f}^a [A]
 \rk \e ^{ - S [a + \cA] + \int \vj \vec{\cA} }\, ,
\ee
which explicitly depends on the background field $a$ since the source couples only to the fluctuation $\cA$ and
since the gauge fixing functional $\tilde{f}^a [A]$ may depend on the background field. The action $S [A] = S [a + \cA]$
is, of course, invariant under both transformations (\ref{ee-9}), (\ref{ee-11}). It is, however, convenient to eliminate
the gauge degrees of freedom by fixing the invariance with respect to the quantum gauge transformation (\ref{ee-9}). 
Then the Faddeev--Popov operator is given by
\be
\label{ee-13}
\tilde{\cM}^{ab} [A] = \frac{\delta_q \tilde{f}^a [A]}{\delta \Theta^b } \, .
\ee
As is well known, the generating functional $\tilde{W} [a, j]$ is invariant under the background gauge transformations
(\ref{ee-11}) if the gauge fixing functional $\tilde{f}^a [A]$ transforms covariantly under these transformations. 
A convenient gauge, which satisfies this requirement, is the background gauge
\be
\label{ee-14}
\tilde{f}^a [A] = [\vd, \vec{\cA}] \equiv [\vd, \vA - \va] = 0 \, ,
\ee
for which the Faddeev--Popov operator (\ref{ee-13}) reads
\be
\label{ee-15}
\tilde{\cM} [A] = - \hat{\vD} \hat{\vd}  \, ,
\ee
where the hat ``$\hat{\phantom{\hspace{0.3cm}}}$'' denotes the adjoint representation of the gauge group
and  $\hat{D}$ is the covariant derivative 
\be
\label{176-ex}
\hat{D}^{ab} = \delta^{ab} \partial + \hat{A}^{ab} , \quad  \hat{A}^{ab}  = \hat{T}^{ab}_c A^c  , \quad \hat{T}^{ab}_c = f^{acb}
\ee
with $f^{abc}$ beeing the structure constants.
This gauge is the $3$-dimensional
spatial analogue of the Landau-deWitt gauge. 
It is not difficult to see that in the
background gauge (\ref{ee-14}) both the gauge constraint and the corresponding Faddeev--Popov determinant are invariant
under the background transformation (\ref{ee-11}). Consequently the generating functional $\tilde{W} [a, j]$
(\ref{ee-12}) and hence also the effective action
\be
\label{ee-16}
\tilde{\Gamma} [a, \bar{\cA}] = \left( \tilde{W} [a, j] - \int \vec{\bar{\cA}} \vec{j}\right)  _{j = j (\bar{\cA})} \,, \quad \bar{\cA} = \frac{\delta \tilde{W} [a, j]}{\delta j}
\ee
are invariant under the background transformation $\delta_b$ (\ref{ee-11}).
Furthermore, from the translation invariance of the integration measure $\cD \cA$ follows that the true
effective action $\Gamma [\bar{A}]$, see Eq.~(\ref{ee-7}), is related to $\tilde{\Gamma} [a, \bar{\cA}]$ by 
\be
\label{ee-18}
\tilde{\Gamma} [a, \bar{\cA}] = \Gamma [a + \bar{\cA}] \, 
\ee
and the classical fields are related by 
\be
\label{ee-19}
\bar{A} = a + \bar{\cA} \, .
\ee   
Obviously the invariance of $\tilde{\Gamma} [a, \bar{\cA}]$ under the background transformation (\ref{ee-11}) ensures
the gauge invariance of $\Gamma [\bar{A}]$.
In this paper we will determine the effective potential $\Gamma [a] = \tilde{\Gamma} [a, \bar{\cA} = 0]$ 
for a constant background field $a$.

\section{\label{III}Hamiltonian approach in the presence of an background field}

Below we develop the Hamilton approach to Yang--Mills theory in the presence of an external background field. This 
approach is an extension of  the Hamiltonian formulation in Coulomb gauge \cite{R22, Feuchter:2004mk} and reduces to the latter when the background
field is switched off. Although in the presence of the background field the Hamiltonian approach could, in principle,
be also formulated in Coulomb gauge the use of the more general background gauge (\ref{ee-14})
turns out to be advantageous.

\subsection{Generalities}

To simplify the bookkeeping we will use the compact notation $A^{a_1}_{k_1} (\vx_1) = A (1)$ for colored Lorentz vectors
like the gauge potential and an analogous notation for colored Lorentz scalars like the ghost $C^{a_1} (\vx_1) = C (1)$. We also define
in coordinate space
\be
\label{aa-172}
\delta (1, 2) = \delta^{a_1 a_2} \delta \lk \vx_1 - \vx_2 \rk \, .
\ee
A repeated label means summation over the discrete color (and Lorentz) indices along with integration over the $d$ spatial coordinates
\be
\label{aa-177}
 A (1) B (1) \equiv \int \d[d] x \sum^d_{k = 1} \sum^{N^2_c - 1}_{a = 1} A^a_k (\vx) B^a_k (\vx) \, .
\ee
Furthermore, indices will be suppressed when they can be easily restored from the context.

The usual canonical quantization of gauge theory assumes Weyl gauge $A_0 = 0$ and results in a 
Hamiltonian
\be
\label{12}
H [A] = \frac{1}{2} \lk g^2 {\vec{\Pi}} ^2 + \frac{1}{g^2}  \vB [A] ^2  \rk
\, ,
\ee
where $\Pi (1) = -\i \delta / \delta A (1)$ is the momentum operator and $B [A]$ is the non-Abelian magnetic
field. Here we have absorbed the coupling constant $g$ in the gauge field, $g A \to A$.
This Hamiltonian is invariant under time-independent gauge transformations. In Weyl gauge Gauss' law escapes the (Heisenberg)
equation of motion and has to be imposed as a constraint on the wave functional
\be
\label{14}
\hat{\vD} \vec{\Pi} \psi [A] = \rho_\text{ext} \psi [A] \, .
\ee 
Here $\rho_\text{ext}$ is the charge density of external matter fields.
Since $\hat{\vD} \vec{\Pi}$ is the generator of (time-independent)
gauge transformations, in the absence of matter fields the wave functional
has to be gauge invariant $\psi [A^U] = \psi [A]$. 

We are interested here in the Yang--Mills vacuum in the presence of an external vector field $a$. Let
$\ket{ \psi_a }$ denote the vacuum wave functional in the presence of the external field $a$ and 
\be
\label{ex-1}
\vev{ \dots }_a \coloneq \bra{\psi_a } \dots \ket{ \psi_a }
\ee
the corresponding vacuum expectation value. We assume here that the states $\ket{\psi_a }$ are
properly normalized, $\bra{ \psi_a } \psi_a \rangle = 1$. In the absence of external fields the gauge invariance of the
vacuum wave functional $\ket{ \psi_{ 0} }$ guarantees that the expectation value of all gauge
dependent quantities vanishes. This refers, in particular, to the expectation value of the gauge field
\be
\label{ex-2}
\bra{ \psi_0 } A \ket{ \psi_0}  =0 \, .
\ee
The background field assigns a prescribed expectation value to the gauge field.
If $\psi_0 [A] = \langle A | \psi_{0} \rangle$ denotes the gauge invariant vacuum wave functional in the
absence of the background field (which necessarily satisfies Eq.~(\ref{ex-2})), then it is trivial to see by a change of variables that the wave functional
\be
\label{ex-3}
\psi_a [A] \coloneq \psi_0 [A - a]
\ee
has the property that 
\be
\label{ex-4}
\vev{ A }_a = a \, .
\ee

Although the wave functional (\ref{ex-3}) satisfies (\ref{ex-4}), in general
 it will not minimize the energy under the constraint (\ref{ex-4}). The wave functional $| \psi_a \rangle$ which minimizes
the energy under the constraint (\ref{ex-4})
\be
\label{603-G9}
\vev{ H }_a \to \text{min}, \quad \vev{ A }_a = a
\ee
can be obtained from the constrained variational principle 
\be
\label{ex-10}
\vev{ H - j(1) A(1) }_a \to \text{extr},
\ee
where $j$ is a Lagrange multiplier chosen such that Eq.~(\ref{ex-4}) is fulfilled.

\subsection{The background gauge fixed Hamiltonian}

Instead of working with gauge invariant wave functionals,
it is usually more convenient to fix the gauge, where those gauges are preferable which allow for an 
explicit resolution of Gauss' law.
The price one pays for the gauge fixing is that the Hamiltonian 
acquires a more complicated, usually non-local form. 
A convenient gauge in this respect is Coulomb gauge \cite{R22}. However, in the present paper, where we are interested in the
energy density of the vacuum in the presence of an external background field $a$, 
Coulomb gauge is not the optimal choice. In this case it is more
convenient to choose the background gauge (\ref{ee-14}) 
which gauge fixes the fluctuating field $\cA = A - a$ with respect to the background field $a$. 

In view of the gauge (\ref{ee-14}) 
it is appropriate to introduce generalized longitudinal and transversal projection operators
\be
\label{18}
\hat{l}_{ij} (x) = \hat{d}_i \lk \hat{\vd} \hat{\vd} \rk^{- 1} \hat{d}_j, \quad \hat{t}_{ij} = 
\hat{\delta}_{ij} - \hat{l}_{ij}, \quad
\hat{\delta}^{ab}_{ij} = \delta^{ab} \delta_{ij} \,  ,
\ee
where
\be
\label{ymt-687}
\hat{d}^{ab}_i = \delta^{ab} \partial_i + \hat{a}^{ab}_i \, , \quad \quad \hat{a}^{ab}_i = f^{acb} a^c_i \, ,
\ee
is the covariant derivative (\ref{ee-15}) with the total gauge field replaced by the background field.
For a constant background field, these projectors have the same 
properties as the ordinary longitudinal and transversal projectors as far as their spatial indices are concerned. However, they are non-trivial matrices in 
color space. 
Using these projectors we split the gauge field and the 
momentum operator into ``longitudinal'' and ``transversal'' parts\footnote{We keep here the terms ``longitudinal''
and ``transversal'' although the components $A^{||}$ and $A^\perp$ have this property only for a vanishing
background  field $a = 0$.}
\bea
\label{19}
A & = & A^{||} + A^\perp, \quad A^{||}_i = \hat{l}_{ij} A_j \nonumber\\
\Pi & = & \Pi^{||} + \Pi^\perp , \quad \Pi^{||}_i = \hat{l}_{ij} \Pi_j \, .
\eea
The longitudinal part of the gauge field $A^{||}$ will be later eliminated by the background 
gauge fixing, Eq.~(\ref{ee-14}), while the
longitudinal part of the momentum operator $\Pi^{||}$ is eliminated by resolving Gauss' law (\ref{14}), which can be explicitly done in the gauge (\ref{ee-14}) 
as we will show now. 

Inserting Eq.~(\ref{19}) into Gauss' law (\ref{14}) 
and solving for the longitudinal part of the momentum operator 
we find
\be
\label{23}
\Pi^{||} \psi = - \hat{d} \lk - \hat{\vD}\hat{\vd} \rk^{- 1} \rho \psi \, ,
\ee
where
\be
\label{aa-274}
\rho = \rho_\text{ext} + \rho_\text{dyn} [A]
\ee
is the total color charge density, which contains besides the external charge $\rho_\text{ext}$ also the dynamical 
charge of the gauge bosons in the background gauge (\ref{ee-14})
\be
\label{24}
\rho_\text{dyn} [A] = - \hat{\vD}\vec{\Pi}^\perp \, .
\ee
Rewriting the covariant derivative as
\be
\label{22}
\hat{D} = \hat{d} + \lk \hat{A} - \hat{a} \rk
\ee
and using $\hat{\vd} \vec{\Pi}^\perp = 0$, the dynamical charge becomes
\be
\label{361}
\rho_\text{dyn} [A] = - \lk \hat{\vA} - \hat{\va} \rk \vec{\Pi}^\perp \, .
\ee
It depends only on the fluctuation $\cA = A - a$ of the gauge field $A$ around the background field $a$. 
Since the $\hat{l}$, $\hat{t}$ (\ref{18}) are orthogonal projectors we have
\be
\label{25}
\vec{\Pi}^2 = \vec{\Pi}^{||} {}^2  + \vec{\Pi}^\perp {}^2 \, .
\ee
Using this relation and
Eq.~(\ref{23}) we find for the gauge fixed Yang--Mills Hamiltonian
\begin{align}
H [A] =& \frac{1}{2} \lk g^2 J^{- 1} [A] {\vec{\Pi}}^\perp  J [A] {\vec{\Pi}}^\perp + \frac{1}{g^2} \vec{B}
 [A^\perp]  \vec{B} [A^\perp] \rk \nonumber \\
&+ H_\text{C} [A] \, ,
\label{26}
\end{align}
where
\be
\label{371}
J [A] = \Det \tilde{\cM} [A] = \Det (-  \hat{\vD} \hat{\vd})
\ee
is the Faddeev--Popov determinant (cf. Eq.~(\ref{ee-15})) and
\be
\label{27}
H_\text{C} [A] = \frac{g^2}{2}  J^{- 1} [A] \rho (1) J [A] F [A] (1, 2) \rho (2) 
\ee
arises from the elimination of the longitudinal component of the momentum operator $\Pi^{||}$ by means of Gauss' law,
see Eq.~(\ref{23}), and describes the interaction between the color charges. Here
\be
\label{28}
F [A] = \lk - \hat{\vD}    \hat{\vd} \rk^{- 1} \lk - \hat{\vd}  \hat{\vd} \rk \lk - \hat{\vD}   \hat{\vd} \rk^{- 1}
\ee
is the analogue of the so-called Coulomb kernel \cite{Feuchter:2004mk} in the background gauge (\ref{ee-14}). 
For a vanishing background
field $a = 0$  the covariant background
derivative $\hat{d} = \partial + \hat{a}$ becomes the $\nabla$-operator
and the kernel (\ref{28}) reduces to the ordinary Coulomb kernel \cite{Feuchter:2004mk}.

In the gauge-fixed theory the matrix elements of an observable $\cO[\Pi, A]$ are defined by
\be
\label{6}
\langle \psi \vert \cO \vert \phi \rangle = \int \cD A \,\delta (\tilde{f} [A]) J [A] \psi^* [A] \, \cO[\Pi, A]\, \phi [A] \, ,
\ee
where $\tilde{f} [A] = 0$ is the gauge fixing constraint (\ref{ee-14}) and $J[A]$ (Eq.~(\ref{371})) the corresponding Faddeev--Popov determinant (\ref{371}).

\subsection{Choice of the wave functional}

In Ref.~\cite{Feuchter:2004mk} a variational determination of the Yang--Mills vacuum wave functional was carried out 
(in the absence of an external
background field) in Coulomb gauge using the Gaussian type wave functional\footnote{To be more precise in Ref.~\cite{Feuchter:2004mk} 
the ansatz (\ref{ex-11}) was used with $J [A]$ (\ref{371}) replaced by the Faddeev--Popov determinant in 
Coulomb gauge $\Det (- \vD \vec{\partial})$. Furthermore in the present case $\omega(1,2)$ can have a non-trivial color structure.}
\bea
\label{ex-11}
\psi_0 [A] & = & J^{- 1/2} [A] \tilde{\psi} [A] \nonumber\\
\tilde{\psi} [A] &\coloneq  & \cN \exp \left[ - \frac{1}{2 g^2} A (1) \omega (1, 2) A (2) \right] \, .
\eea
Here $\cN$ is a normalization constant and $\omega$ is the variational kernel. 
The ansatz (\ref{ex-11}) has the advantage that it removes the Faddeev--Popov determinant $J [A]$ 
from the integration measure (\ref{6}).
Here we extend the 
variational calculations to the presence of an external background field $a$. To
fullfill the constraint $\vev {A} _a = a$  we use the 
trial wave functional
\be
\label{ex-12}
\psi_a [A] = J^{- 1/2} [A] \tilde{\psi} [A - a] \, ,
\ee
with $\tilde{\psi} [A]$ defined in Eq.~(\ref{ex-11}). In
the absence of the background field $\psi_a [A]$ reduces to the vacuum wave functional $\psi_0 [A]$ (\ref{ex-11}). 
This trial wave functional satisfies already the constraint
(\ref{ex-4}), so it remains to minimize the energy 
$\vev { H }_a$ with respect to the kernel $\omega (1, 2)$.
Due to the presence of the colored background field
the resulting kernel $\omega (1, 2)$ will be a non-trivial matrix in color space.

With the trial wave functional (\ref{ex-12}) we find for the expectation value 
of any observable $\cO [\Pi,
A ]$ from Eq.~(\ref{6})
after the shift\footnote{Note 
that the momentum operator $\Pi$ remains unchanged under the shift of coordinates $(A - a) \to A$.} $(A - a) \to A$ of integration variables 
\bea
\label{ex-13}
\vev { \cO }_a & \equiv & \langle \psi_a | \cO [\Pi, A] | \psi_a \rangle \\
& = & \int \cD A \,\delta (\tilde{f} [A + a]) \tilde{\psi}^* [A] \,\tilde{\cO} 
\left[\Pi, A + a \right]\, \tilde{\psi} [A]\nonumber \\ 
&\eqcolon &\vev { \cO \left[ \Pi, A + a \right] }_{0}  \nonumber \, ,
\eea
where we have introduced the abbreviation
\be
\label{ex-14}
\tilde{\cO} [\Pi, A] = J^{1/2} [A] \cO [\Pi, A] J^{- 1/2} [A]  \, .
\ee
Equation (\ref{ex-13}) is the vacuum expectation value of the observable $\tilde{\cO} [\Pi, A + a]$ in the gauge fixed theory with the gauge
condition, Eq.~(\ref{ee-14})
\be
\label{ex-7}
\tilde{f} [A + a] = \hat{\vd} \vA = 0 \, .
\ee
Note that in $\vev{ \cO }_a$, (\ref{ex-13}), the gauge constraint (\ref{ex-7}) is implemented, which eliminates
the longitudinal gauge field $A^{||}$ defined by Eq.~(\ref{19}). Equation (\ref{ex-13}) applies in particular to the 
gauge fixed Hamiltonian (\ref{26}). Note also that after the shift of variables $(A- a) \to A$ the color charge of the gauge field (\ref{361}) becomes 
\be
\label{ex-15}
\rho_\text{dyn} [A + a] = - \hat{\vA} \vec{\Pi}^\perp \, ,
\ee 
which is formally the same expression as obtained in Coulomb gauge \cite{R22, Feuchter:2004mk} except that transversality is now 
defined by the gauge condition (\ref{ex-7}) and thus depends on the background field $a$.

In passing we notice that in the gauge fixed Hamiltonian $H [\Pi, A]$ (\ref{26})
the gauge field enters only
in form of the covariant derivative $\hat{D} = \partial + A$.  
(Recall that the non-Abelian magnetic 
field can be written as $B_i = \frac{1}{2} \epsilon_{ijk} [ \hat{D}_j, \hat{D}_k ]$.) Therefore, after
the shift of variables $A \to A + a$ the Hamiltonian $H [\Pi, A + a]$ depends on the background field 
$a$ only in the combination with the $\nabla$-operator $\partial + \hat{a} = \hat{d}$. This is, in particular, true for the
Faddeev--Popov determinant (\ref{ee-15})
\be
\label{858-11}
J [A + a] = \Det \lk - \lk \hat{\vd} + \hat{\vA} \rk \hat{\vd} \rk 
\ee
and thus also for the transformed Hamiltonian $\tilde{H} [\Pi, A + a]$ defined by 
Eq.~(\ref{ex-14}).  

We are eventually interested in the effective potential of the order parameter of confinement. For this purpose it is sufficient
to consider constant background fields $\va$, which we will assume from now on.
Note, since $\hat{\va} \cdot \va = 0$, 
for constant background fields $\va$ we have $\hat{\vd} \cdot \va = 0$ and the gauge condition 
(\ref{ee-14}) reduces to the 
condition (\ref{ex-7}) $\hat{\vd} \cdot \vA = 0$. For the time being the color structure of
the background field $\hat{a} = a^b (\vx) \hat{T}_b$ is arbitrary. 

\section{\label{IV}Yang--Mills dynamics in the presence of a background field}

The presence of the background field influences the form of the propagators. In addition, the 
propagators depend on the gauge chosen. In the following we present the equation of motion for the ghost and 
gluon propagators following from the constrained variational principle (\ref{603-G9}) in the background gauge (\ref{ex-7}) with the trial wave functional (\ref{ex-12}). Thereby
we will make no assumption on the form of background field (except that it is constant). 
The resulting equations of motion are direct generalizations of the
equations obtained in the variational approach in Coulomb gauge \cite{Feuchter:2004mk} and reduce to the latter for a vanishing
background field.

\subsection{The gluon propagator}

Since by Eq.~(\ref{ex-13}) all expectation values $\langle \ldots \rangle_a$ in the state $| \psi_a \rangle$, satisfying the constraint (\ref{ex-4}), 
can be reduced to the vacuum expectation value $\langle \ldots \rangle_{0}$ it suffices to consider
the gluon propagator 
\be
\label{495-mm}
\cD (1, 2) = \langle A (1) A (2) \rangle_0 \, ,
\ee
which for the trial wave functional (\ref{ex-11}) is given by
\be
\label{500-mm}
\cD (1, 2) = \frac{g^2}{2} \omega^{- 1} (1, 2) \, .
\ee
Independent of the form of the background field from the definition (\ref{495-mm}) follows the symmetry relation
$\cD (1, 2) = \cD (2, 1)$,
i.e.
\be
\label{dd-849-x2}
\cD^{ab}_{kl} (\vx, \vy) = \cD^{ba}_{lk} (\vy, \vx) \, .
\ee
Since the gauge field is ``transverse'' (w.r.t. the covariant background derivative $d$ (\ref{ymt-687}))
\be
\label{dd-854-x3}
A^a_k (\vx) = t^{ab}_{kl} (\vx) A^b_l (\vx)
\ee
the gluon propagator is also ``transverse'' satisfying
\be
\label{dd-859-x4}
t^{aa'}_{kk'} (\vx) t^{bb'}_{ll'} (\vy) \cD^{a' b'}_{k' l'} (\vx, \vy) = 
\cD^{ab}_{kl} (\vx, \vy) \, .
\ee
Note, the ordering is here important since $t (\vx)$ is a differential operator. By Wick's theorem expectation values in the state $\psi_0 [A]$ (\ref{ex-11})
can be entirely expressed in terms of the gluon propagator (\ref{495-mm}), see Sect.~\ref{sectD}.

\subsection{The ghost DSE}

In the presence of the external background field the ghost propagator is defined by
\be
\label{aa-493}
G = \langle \tilde{\cM}^{- 1} [A] \rangle_a \, ,
\ee
where $\tilde{\cM} [A]$ is the Faddeev--Popov kernel (\ref{ee-15}) and $\langle \dots \rangle_a$ is defined in 
Eq.~(\ref{ex-13}), from which follows that the ghost propagator can be expressed as vacuum expectation value
\be
\label{aa-499-x1}
G = \vev{ \tilde{\cM}^{- 1} \left[ A + a \right] }_0 = \vev{ \lk - \lk \hat{\vd} + \hat{\vA} \rk  \hat{\vd} \rk^{- 1} }_0\, .
\ee
In the way described in Ref.~\cite{Feuchter:2004mk} one derives for this ghost propagator the following Dyson--Schwinger equation 
\be
\label{aa-504-x2}
G^{- 1} = G^{- 1}_0 - \Sigma \, ,
\ee
where $G_0$ is the bare ghost propagator, which is defined by Eq.~(\ref{aa-499-x1}) with $A = 0$
\be
\label{aa-509-x3}
G_0 = \lk - \hat{\vd}  \hat{\vd} \rk^{- 1} \, .
\ee
Furthermore, the ghost self-energy is given by
\be
\label{aa-514-x4}
\Sigma (1, 2) = \Gamma_0 (1, 3; 4) G (3, 3') \Gamma (3', 2; 4') \cD (4', 4) \, ,
\ee
where $\cD (1, 2)$ is the gluon propagator (\ref{495-mm}) and
\be
\label{aa-519-x5}
\Gamma_0 (1, 2; 3) = \frac{\delta \tilde{\cM} \left[ A + a \right] (1, 2)}{\delta A (3)} 
\ee
is the bare ghost-gluon vertex. The full ghost-gluon vertex $\Gamma$ is defined here by
\be
\label{aa-524}
\vev{ \tilde{\cM}^{- 1} \left[ A + a \right] \Gamma_0 \tilde{\cM}^{- 1} \left[ A + a \right] }_0 = G \Gamma G \, .
\ee
Following Ref.~\cite{Feuchter:2004mk} we use the rainbow ladder
approximation replacing the full ghost-gluon vertex by the bare one. The justification for this is the following:
As shown in Landau gauge \cite{R24} the ghost-gluon vertex is not renormalized. This holds also in Coulomb gauge \cite{Schleifenbaum:2006bq}. 
Furthermore, in Coulomb gauge the dressing of the ghost-gluon vertex, at least to one-loop order, is small \cite{Campagnari:2011bk}.
We assume that this remains valid for the present gauge (\ref{ex-7}), which reduces to Coulomb gauge in the 
absence of the background field.

With the explicit form of the Faddeev--Popov kernel $\tilde{\cM} [A]$ (\ref{ee-15})
we find for the bare ghost-gluon vertex (\ref{aa-519-x5}) 
\begin{multline}
\label{y23}
{\Gamma}^a_{0, k} (\vx_1, \vx_2; \vx_3) \\
= \lk \hat{T}_b t^{ba}_{lk} (\vx_1) \delta (\vx_1 - \vx_3) \rk \hat{d}_l (\vx_1)
\delta (\vx_1 - \vx_2) \, ,  
\end{multline}
with $\hat{T}^{ac}_b = f^{abc}$. Here we have suppressed the adjoint color indices $a_1, a_2$, of the ghost legs and set
\be
\label{aa-545}
\Gamma (1, 2; 3) \equiv \lk \Gamma_{a_3 k_3} \lk \vx_1, \vx_2; \vx_3 \rk \rk^{a_1 a_2} \, .
\ee
Replacing the full ghost-gluon vertex $\Gamma$ by the bare one (\ref{y23}) we find for the ghost self-energy (\ref{aa-514-x4})
\begin{multline}
\label{y-782}
{\Sigma} (\vx_1, \vx_2)  = \left[ t^{aa'}_{kk'} (\vx_1) t^{bb'}_{ll'} (\vx_3) D^{b' a'}_{l' k'} \lk \vx_3, \vx_1 \rk \right] 
\\
 \times \hat{T}_a \lk \hat{d}_k (\vx_1) {G} (\vx_1, \vx_3) \rk \hat{T}_b \lk \hat{d}_l (\vx_3) \delta \lk \vx_3, \vx_2 \rk \rk \, ,
\end{multline}
which is still a matrix in adjoint color space.

\subsection{\label{subC}The curvature}

The kinetic part of the transformed Yang--Mills Hamiltonian $\tilde{H}$ cf. (\ref{ex-14}) contains functional derivatives of the
Faddeev--Popov determinant. In the vacuum expectation value of $\tilde{H}$ (see  subsection \ref{sectD}) 
these derivatives enter in form of the ghost loop
\be
\label{y29}
\chi (1, 2) = - \frac{1}{2} \vev{ \frac{\delta^2 \ln J
 [A + a]}{\delta A (1) \delta A (2)}  }_{0} \, ,
\ee
which in the present context is referred to as curvature \cite{Feuchter:2004mk}. 
The curvature is defined here as in Ref.~\cite{Feuchter:2004mk}, however, with the argument of the Faddeev--Popov determinant shifted
by the background field.
With the definition of the ghost-gluon vertex (\ref{aa-524}) it can be expressed as (cf. Ref.~\cite{Feuchter:2004mk} for more details)
\be
\label{y30}
\chi (1, 2) = \frac{1}{2} \Tr \lk G \Gamma (1) G \Gamma_0 (2) \rk \, .
\ee
Using again the bare ghost-gluon vertex approximation one finds with (\ref{y23}) 
\begin{multline}
\label{yy-886}
\chi^{ab}_{kl} (\vy_1, \vy_2) = \, \frac{1}{2} t^{ aa'}_{kk'} (\vy_1) t^{bb'}_{ll'} (\vy_2) \\
\times \tr \left[ \hat{T}_{a'} \lk \hat{d}_{k'} (\vy_1) G (\vy_1, \vy_2) \rk \hat{T}_{b'} \hat{d}_{l'} (\vy_2) G (\vy_2, \vy_1) \right] \, ,
\end{multline}
where the trace is over the adjoint color space.

\subsection{\label{sectD}The energy density and the gap equation}

Since the vacuum wave functional $\tilde{\psi} [A]$ (\ref{ex-11}) is Gaussian it is straightforward to calculate the 
energy  $\vev{ H }_a$ (cf. (\ref{ex-13})) by using
Wick's theorem and expressing $\vev { H \left[\Pi, A + a \right] }_0$ as a functional of the gluon propagator $\cD$ (\ref{495-mm}). 
For the Abelian part of the magnetic energy one finds
\be
\label{618-26}
\vev { H^A_B }_a = \frac{1}{2 g^2} \left[ \lk - \hat{\vd}  \hat{\vd} \rk (1, 2) \cD (2, 1') \right]_{1' = 1}
\ee
while the non-Abelian part gives 
\begin{multline}
\label{1447-ll}
\vev { H^{NA}_B }_a = \, \frac{1}{4 g^2} f^{abc} f^{a b' c'}  \int \d[d] x \Big[ \cD^{bb'}_{ll} (x, x) \cD^{cc'}_{mm} (x, x) \\
+ \cD^{bc}_{lm} (x, x)  \cD^{b' c'}_{lm} (x, x) 
+ \cD^{b c'}_{lm} (x, x) \cD^{b' c}_{lm} (x, x) \Big] \, .
\end{multline}
For the expectation value of the kinetic part $\tilde{H}_K$ of the transformed Hamiltonian (\ref{ex-14})

\begin{multline}
\tilde{H}_K [A] = \frac{g^2}{2} \bigg\{ \Pi (1) \Pi (1) + \frac{1}{2} \lk 
\frac{\delta}{\delta A (1)} \frac{\delta}{\delta A (1)}
\ln J [A] \rk\\
 + \frac{1}{4} \frac{\delta \ln J [A]}{\delta A (1)} \frac{\delta \ln J [A]}{\delta A (1)} \bigg\} 
\end{multline}
one finds, using the definition of the curvature (\ref{y29}) (see Refs.~\cite{Feuchter:2004mk, R16} for more details)
\begin{multline}
\label{636-28}
\vev{ H_K }_a \\= \frac{g^2}{2} \left\{\frac{\cD^{- 1} (1, 1)}{4}  - \chi (1, 1) + \chi (1, 2) \cD (2, 3) \chi (3, 1) \right\} \, .
\end{multline}
Variation of the energy $\vev { H }_a$ with respect to the gluon propagator $\cD$ (\ref{495-mm})
yields the gap equation.
As shown in Ref.~\cite{R16} the Coulomb term $H_\text{C}$ has little influence on the gluon sector. Furthermore,
 the non-Abelian part of the magnetic energy
gives rise to a tadpole, which in the absence of the background field
 contributes a (UV-diverging) constant to the gap equation, which can be absorbed into a renormalization
constant. Ignoring the Coulomb term and the non-Abelian part of the magnetic energy the gap equation arising from
\be
\label{844-ae}
\vev { H_K + H^A_B }_a \to \min \,
\ee
 becomes 
\be
\label{643-29}
\frac{g^4}{4}  \cD^{- 1} (2, 3) \cD^{- 1} (3, 1) = (- \hat{t} \hat{\vd}  \hat{\vd} \hat{t}) 
(2, 1) + g^4 \chi (2, 3) \chi (3, 1) \, .
\ee
Here we have also ignored the derivatives $\delta \chi / \delta \cD$, which
give rise to two-loop terms since $\chi$ itself contains already one ghost loop.
Using the gap equation (\ref{643-29}) to express $(- \hat{\vd}  \hat{\vd})$ in the magnetic energy in terms of $\cD$ and $\chi$
the vacuum energy can be cast into the compact form 
\be
\label{650-30}
\langle H_K + H^A_B \rangle_a = g^2 \left[ \frac{1}{2} \cD^{- 1} (1, 1) - \chi (1, 1) \right] \, .
\ee
In Sect.~\ref{sectVI} we will extract from this expression the effective potential of the order parameter of confinement.

\section{\label{V}The propagators in the presence of the background field}

The above considerations are valid for any gauge group and
for arbitrary constant background fields. For pedagogical reason we will confine ourselves below to the gauge group SU$(2)$. The extension to SU$(N)$ 
is straightforward and will be presented in Subsect.~\ref{sectVI-neu}.

\subsection{Choice of the background field}

As shown in Sect.~\ref{sectionII-neu} to find the effective potential of the confinement order parameter, it is sufficient to 
consider constant background fields $\va = \text{const}$. 
By isotropy of space, without loss of generality we can direct
the background field along the $3$-axis
\be
\label{aa-835}
{\va} = {a} \ve_3 \, .
\ee
Furthermore, as we have seen in Sect.~\ref{sectionII-neu} the background field has to live in the Cartan subalgebra to figure as order parameter of 
confinement. Therefore we have to choose for SU($2$)
\be
\label{aa-847}
{a}  \equiv {a}^b {T}_b = a {T}_3 \, , \quad a = \text{const}\, .
\ee
In the Hamiltonian approach presented above the
 background field occurs in the adjoint representation. The Cartan generator in the adjoint representation $\hat{T}_3 = \epsilon^{a 3 b}$
 is diagonalized in the basis of the spin $s = 1$
eigenstates
\be
\label{455-17}
\hat{T}_3 \ket{\sigma } = - \i \sigma | \sigma \rangle , \quad \sigma = 0, \pm 1 \, ,
\ee
where, in the usual bracket notation, the
\be
\label{460-*18}
\langle a | \sigma \rangle =  e^a_\sigma
\ee
are the cartesian components of the spherical unit vectors (in color space)
\be
\label{y5}
\ve_{\sigma = 1} = - \frac{1}{\sqrt{2}}  \begin{pmatrix} 1\\ \i \\ 0 \end{pmatrix}, \,
\ve_{\sigma = - 1} = \frac{1}{\sqrt{2}}  \begin{pmatrix} 1\\ - \i \\ 0 \end{pmatrix} ,\, 
\ve_{\sigma = 0} =  \begin{pmatrix} 0\\ 0 \\ 1 \end{pmatrix} .
\ee
Here and in the following Greek letters $\sigma, \tau, \rho, \dots$ denote spherical color components $(1, 0, - 1)$
while Latin letters $a, b, c, \dots$ denote the Cartesian color components $(1, 2, 3)$. 
The vectors $\ve_\sigma$ satisfy the symmetry relation 
\be
\label{477-19}
\ve^*_\sigma = (-)^\sigma \ve_{- \sigma}
\ee
and form an orthonormal 3-bein
\be
\label{482-20}
\ve^*_\sigma  \ve_\tau = \delta_{\sigma \tau}, \quad \ve_\rho \times \ve_\sigma = - \i \epsilon_{\rho \sigma \tau}
\ve^*_\tau \, .
\ee
By Eq.~(\ref{455-17}) we have
\be
\label{1123-x}
\hat{a} | \sigma \rangle = a \hat{T}_3 | \sigma \rangle = - i a \sigma | \sigma \rangle \, .
\ee
Any matrix in the adjoint representation can be expressed in the spherical basis as
\be
\label{488-21}
M^{ab} = \sum_{\sigma, \tau} \langle a | \sigma \rangle M^{\sigma \tau} \langle \tau | b \rangle \, .
\ee
Due to Eq.~(\ref{1123-x}), the covariant derivative (\ref{ymt-687}) $\hat{d}_k$ becomes diagonal in the spherical basis 
\be
\label{493-22}
d^{\sigma \tau}_k = \delta^{\sigma \tau} d^\sigma_k, \quad d^\sigma_k = \partial_k + \delta_{k3} (- \i \sigma a) \, .
\ee
The same is true for the longitudinal and transversal projectors (\ref{18})
\bea
\label{498-23}
l^{\sigma \tau}_{kl} & = & 
\delta^{\sigma \tau} l^\sigma_{kl}, \quad l^\sigma_{kl} = d^\sigma_k \lk \vd^\sigma \vd^\sigma \rk^{- 1}
d^\sigma_l \nonumber\\
t^{\sigma \tau}_{k l} & = & \delta^{\sigma \tau} t^\sigma_{kl}, \quad t^\sigma_{kl} = \delta_{kl} - l^\sigma_{kl} 
\, ,
\eea
which makes the spherical basis favorable. 
Defining the momentum representation of these quantities by
\bea
\label{aa-732}
d (x) e^{\i \vp \vx} & = & d (\vp) e^{\i \vp \vx} \nonumber\\
t (x) e^{\i \vp \vx} & = & t (\vp) e^{\i \vp \vx}
\eea
we have in the spherical basis
\bea
\label{aa-903-G14}
d^\sigma_k (\vp) &= & \i \lk \vp - \sigma a \ve_3 \rk_k = \i \lk p_k - \sigma a \delta_{k 3} \rk \, , \\
\label{aa-903}
t^\sigma_{kl} (\vp) &= & \delta_{kl} - \frac{d^\sigma_k (\vp) d^\sigma_l (\vp)}{\vd^\sigma (\vp) \vd^\sigma (\vp)} \, ,
\eea
where
\be
\label{aa-909-G9}
- \vd^\sigma (\vp) \vd^\sigma (\vp) = \lk \vp - \sigma a \ve_3 \rk^2 = \vp^2_\perp + \lk p_3 - \sigma a \rk^2 
\ee
and $\vp_\perp$ is the projection of $\vp$ on the 1-2-plane. Furthermore these quantities satisfy the symmetry relations
\begin{align}
t^\sigma_{kl} (\vp) &= t^\sigma_{lk} (\vp)  \,,\quad t^\sigma_{kl} (- \vp) = t^{- \sigma}_{kl} (\vp) \,,\nonumber \\
d^\sigma_k (- \vp) &= - d^{- \sigma}_k (\vp)  \,.
\label{aa-914}
\end{align}

\subsection{Propagators}

For a constant background field the homogeneity of space is preserved and the Green's functions can depend 
only on coordinate differences and thus can be Fourier transformed as in the absence of the background field. 
We define the Fourier transform for the gluon propagator by
\be
\label{aa-823}
\cD (\vx, \vy) = \int \dbar p\, \e^{\i \vp (\vx- \vy)} \cD (\vp), \quad \dbar[] p \equiv \frac{\d\, p}{(2 \pi)^d}
\ee
and analogously for the other two-point functions like ghost propagator and curvature.
The background field (\ref{aa-847}), however, singles out a direction in $3$-space and thus spoils SO$(3)$ invariance.
Therefore the Fourier transformed two-point functions, like $\cD (\vp)$, will not just depend on the modulus
$\abs{\vp}$ but also on the direction of $\vp$. 

Since the background field $\hat{a} = a \hat{T}_3$ is diagonal in the spherical basis and since $\hat{a}$
is the only color dependent quantity we expect that the various two-point functions like 
gluon and ghost propagators are also diagonal in this basis. We show in appendix \ref{appA} that the equations of motion 
can indeed be consistently solved for color diagonal propagators of the form
\begin{align}
\label{aa-852-z1}
\cD^{\sigma \tau}_{kl} (\vp) &= \delta^{\sigma \tau}  t^\sigma_{kl} (\vp) \cD^\sigma (\vp) \\
\label{aa-855-z2}
G^{\sigma \tau} (\vp) &= \delta^{\sigma \tau} G^\sigma (\vp) \, .
\end{align}
For the reduced propagators $G^\sigma (p)$, $\cD^\sigma (p)$ the equations of motion simplify drastically. From the appendix~\ref{appA}  
we find for the ghost DSE (\ref{930-x4})
\be
\label{1183-x1}
G^\sigma (\vp)^{- 1} = G^\sigma_0 (\vp)^{- 1} - \Sigma^\sigma (\vp) \, ,
\ee
where
\be
\label{1188}
G^\sigma_0 (\vp)^{- 1} = - \vd^\sigma (\vp) \vd^\sigma (\vp) = \lk \vp - \ve_3 \sigma a \rk^2 
\ee
is the bare (inverse) ghost propagator (\ref{aa-509-x3}) and 
\begin{multline}
\label{1193-x2}
\Sigma^\sigma (\vp) =\\
 - \sum_\mu \int \dbar  q \,d^\sigma_l (\vp) t^\mu_{lk} (\vq) d^\sigma_k (\vp) \cD^\mu (\vq) G^{\sigma + \mu}
(\vp + \vq)
\end{multline}
is the ghost self-energy (\ref{y-782}). Here and in the following sums of spherical indices, like $\mu + \sigma$, are defined modulo $3$. As shown in the appendix~\ref{appA} we find for the gap equation (\ref{1015-y1}) after the transversal projectors are 
contracted
\be
\label{1200-x3}
\frac{g^4}{4} \lk \cD^\sigma (\vp) \rk^{- 2} = - \vd^\sigma (\vp) \vd^\sigma (\vp) +
g^4 \lk \chi^\sigma (\vp) \rk^2 \, ,
\ee
where (cf. Eq.~(\ref{1266-ff-x1}))
\begin{multline}
\label{1206-x4}
\chi^\sigma (\vp) = - \frac{g^2}{2 (d - 1)}\\
\times \sum_\mu \int  \dbar[] \,q d^\mu_k (\vq) g^\sigma_{kl}
(\vp) d^\mu_l (\vq) G^\mu (\vq) G^{\sigma + \mu} (\vp + \vq)
\end{multline}
is the scalar curvature. Finally from Eq.~(\ref{1241-qq}) we have for the energy (\ref{650-30})
\begin{multline}
\label{1212-18}
\langle H_K + H^A_B \rangle_a = \\
(d - 1) V \sum_\sigma \int \dbar q \,\lk \frac{g^2}{2}\cD^\sigma (\vq)^{- 1} - \chi^\sigma (\vq) \rk .
\end{multline}
Let us emphasize that $G^\sigma (\vp)$, $\Sigma^\sigma (\vp)$, $\cD^\sigma (\vp)$ and
$\chi^\sigma (\vp)$ are all scalars in $\mR^3$ but have a non-trivial color structure due to their dependence on 
the spherical color index $\sigma$. 

\subsection{\label{VF}Relating the propagators in the presence of the background field to the vacuum propagators in Coulomb gauge}

The solutions of the constraint variational problem (\ref{603-G9})
are defined by the coupled equations (\ref{1183-x1}), (\ref{1200-x3}), in which the expressions (\ref{1193-x2}) 
$\Sigma^\sigma (\vp)$ and (\ref{1206-x4}) $\chi^\sigma (\vp)$ enter.
It is not difficult to see that solutions exist, which obey the symmetry relations
\be
\label{1112-x1}
\cD^\mu (- \vp) = \cD^{- \mu} (\vp) \,,\quad G^\mu (- \vp) = G^{- \mu } (\vp) \, .
\ee
Indeed assuming that these relations hold, one finds from the explicit expressions for $\Sigma$ (\ref{1193-x2}) and $\chi$ (\ref{1206-x4}) by using (\ref{aa-914})
\be
\label{1118-oo}
\Sigma^\mu (- \vp) = \Sigma^{- \mu} (\vp) \, , \quad \chi^\mu (- \vp) = \chi^{- \mu} (\vp) \, ,
\ee
which in turn entails Eqs. (\ref{1112-x1}). 
We will now show that these solutions
can be related to the vacuum solution in Coulomb gauge obtained in Ref.~\cite{Feuchter:2004mk}. 

Fourier transforming Eq.~(\ref{500-mm}), going to the spherical basis (\ref{488-21}) and assuming (\ref{aa-852-z1}) we have
\be
\label{1026-oo}
\cD^\sigma (\vp) = \frac{g^2}{2 \omega^\sigma (\vp)} \, ,
\ee
where $\omega^\sigma (\vp)$ is the energy of a gluon with spherical color 
component $\sigma$. 
Using also Eqs.~(\ref{1188}) and Eq.~(\ref{1240-ff}) from appendix \ref{appA} the gap equation (\ref{1200-x3}) becomes
\be
\label{1031-z1}
\lk \omega^\sigma (\vp) \rk^2 = \lk \vp - \sigma a \ve_3 \rk^2 + \lk \chi^\sigma (\vp) \rk^2  \, ,
\ee
which is now a scalar equation. It differs from the vacuum gap equation in Coulomb gauge obtained in Ref.~\cite{Feuchter:2004mk}  by the
explicit presence of background field $a$ in the first term of the r.h.s.\footnote{As,
compared to Ref.~\cite{Feuchter:2004mk}, in the present paper we have ignored the tadpole and the Coulomb term.} and by the index $\sigma$ labeling the eigenvalues
of $- \i \hat{T}_3$. Obviously, 
the gap equation (\ref{1031-z1}) allows for solution of the form
\be
\label{1043-oo}
\omega^\sigma (\vp) = \omega (\vp_\sigma) \, ,
\ee
where
\be
\label{1048-oo}
\vp_\sigma \coloneq \vp - \sigma a \ve_3
\ee
and $\omega (\vp)$ is the solution in the absence of the background field, i.e. for $a = 0$, provided also the curvature has 
the property:
\be
\label{1054*oo}
\chi^\sigma (\vp) = \chi (\vp_\sigma) \, .
\ee
We will now explicitly show that both the gap equation (\ref{1031-z1}) and the ghost DSE (\ref{1183-x1}) are indeed satisfied  by
propagators of the form
\be
\label{1060-z2}
\cD^\sigma (\vp) = \cD (\vp_\sigma) \, , \quad G^\sigma (\vp) = G (\vp_\sigma) \, ,
\ee
where $\cD (\vp)$ and $G (\vp)$ are the corresponding propagators for $a = 0$, i.e. the vacuum solutions of the
Hamiltonian approach in Coulomb gauge \cite{Feuchter:2004mk}. 

First notice that by the definitions (\ref{aa-903-G14}), (\ref{aa-903}) we have
\be
\label{1070-z3}
d^\sigma (\vp) = d (\vp_\sigma) \, , \quad t^\sigma (\vp) = t (\vp_\sigma) \, .
\ee
Assuming (\ref{1060-z2}) and noticing that by Eq.~(\ref{1048-oo})
\be
\label{1075-oo}
(\vp + \vq)_{\sigma + \mu} = \vp_\sigma + \vq_\mu
\ee
from Eq.~(\ref{1193-x2}) follows indeed with (\ref{1070-z3})
\be
\label{1080*z4}
\Sigma^\sigma (\vp) = \Sigma (\vp_\sigma)
\ee
and thus the ghost DSE (\ref{1183-x1}) allows indeed for solutions of the form (\ref{1060-z2}). In the same way
one shows that, assuming (\ref{1060-z2}), from Eq.~(\ref{1206-x4}) follows
\be
\label{1086-oo}
\chi^\sigma (\vp) = \chi (\vp_\sigma) \, ,
\ee
which implies that also the gap equation (\ref{1031-z1}) allows for solutions of the form (\ref{1060-z2}). This completes
the proof of existence of solutions of the form (\ref{1060-z2}). Since the propagators in the absence of the background field
depend only on $\vp^2$ the solutions (\ref{1060-z2}) also satisfy the symmetry relation (\ref{1112-x1}).
We have thus shown that the propagators in the background gauge (\ref{ex-7}) in the state of minimal energy under the
constraint (\ref{ex-4}) are related by Eq.~(\ref{1060-z2}) to the ordinary vacuum propagators in Coulomb gauge\footnote{In principle, there could, of course, be other solutions which do not have the form (\ref{1060-z2}). So far we have no indication that such solutions do exist.}.
The latter have been determined previously in a variational approach in the continuum theory \cite{Feuchter:2004mk, Epple:2006hv} as well as measured
on the lattice \cite{R25}. In the section \ref{sectVI}, we will use these propagators as input to calculate the effective potential
of the confinement order parameter. For this purpose we briefly review in Sect.~\ref{VG} the results obtained for $\omega (\vp)$ and $\chi (\vp)$ in the Hamiltonian approach in Coulomb gauge
at zero temperature.

\subsection{\label{sectVI-neu}Extension to SU($N$)}

The previous considerations can be straightforwardly extended to an arbitrary gauge group SU($N$). As explained in Sect.~\ref{sectionII-neu} the background field $a$ has to be chosen in the Cartan subalgebra
\be
\label{1301-x1}
a = \sum^r_{k = 1} a^k H_k \, ,
\ee
where
\be
\label{1306-x}
H_k\, , \quad k = 1, \dots, r
\ee
are the generators of the Cartan subalgebra with $r$ being the rank of the group. Since the $H_k$ commute with each other they can be 
simultaneously diagonalized. Their eigenvalues are the so-called weights $-\i \mu_k$, which are the entries of the weight vectors
\be
\label{1347-17}
\vmu = \lk \mu_1, \ldots, \mu_r \rk \,  .
\ee
 In the present approach we need the background field in the
adjoint representation $\hat{a}$. The weights in the adjoint representation $\hat{H}_k$ are the roots $\sigma_k$
\be
\label{1313-x2}
\hat{H}_k \ket{ \sigma } = -\i \sigma_k \ket{ \sigma } \, ,
\ee
which are the entries of the root vectors
\be
\label{1318-x}
\vec{\sigma} = \lk \sigma_1, \dots, \sigma_r \rk \, .
\ee
For the gauge group SU($2$), which has rank $r = 1$, the single generator of the Cartan subalgebra is given by $H_1 = T_3$ and the
eigenbasis of $\hat{T}_3$ is explicitly given in Eq.~(\ref{y5}). As in the SU($2$) case, in the Cartan basis (\ref{1313-x2}) the
background field (\ref{1301-x1}) is obviously diagonal
\be
\label{1325-x}
\hat{a} | \sigma \rangle = -\i \va  \vec{\sigma} | \sigma \rangle \, ,
\ee
where $\va = (a_1,\ldots,a_r)$ and
\be
\label{1330-x}
\va \vec{\sigma} = \sum^r_{k = 1} a_k \sigma_k
\ee
is the scalar product in the Cartan subalgebra. Furthermore for SU($N$) the non-vanishing roots come in pairs $\pm \vec{\sigma}$.
 (Some of the roots may vanish as in 
the SU($2$) case. In the mathematical literature
the term ``root'' is usually reserved to the non-vanishing roots.) Besides SU($2$), we will be mainly interested in the gauge group SU($3$),
which has rank $2$. The two generators of the Cartan subalgebra are usually chosen as $H_1 = T_3$ and $H_2 = T_8$. 
For this group the  positive weights read
\be
\label{1462-G19}
\vec{\mu} = \lk 0, \, \frac{1}{\sqrt{3}} \rk  , \quad  \lk \frac{1}{2} , \, \frac{1}{2 \sqrt{3}} \rk  , \quad 
\lk \frac{1}{2}  , \, - \frac{1}{2 \sqrt{3}} \rk
\ee
and the
non-vanishing positive\footnote{A weight or a root is called ``positive''  if its first component is positive.} roots are given by
\be
\label{1338-x}
\vec{\sigma} = (1, 0)   , \quad  \lk \frac{1}{2} , \frac{1}{2}  \sqrt{3} \rk  , \quad \lk \frac{1}{2} , -  \frac{1}{2}  \sqrt{3} \rk \, ,
\ee
where the first root occurs also in the SU($2$) case. 
Note that for SU($3$) the scalar product of a root vector with a weight vector is either $0$ or $\frac{1}{2}$
\be
\label{1368-20}
\vsigma \vmu \in \{ 0\,,\;  \frac{1}{2}\} \, .
\ee
All considerations of the previous section remain valid for SU($3$) with the only modification that the shifted momentum
(\ref{1048-oo}) has to be replaced by
\be
\label{1344-G3}
\vp_\sigma = \vp - (\vec{\sigma} \va) \ve_3 \, ,
\ee
where $\vec{\sigma} \va$ is defined in Eq.~(\ref{1330-x}). With this modification all equations of the previous section remain valid when the
index $\sigma = 0$, $\pm 1$ is replaced in SU($3$) by all ($2$-dimensional) root vectors $\vec{\sigma}$. Since the roots
(\ref{1338-x}) are normalized $\vec{\sigma} \vec{\sigma} = 1$ the co-weight vectors $\tilde{\mu}_k$ (Eq.~(\ref{209-a4})) are given
by $2 \vec{\mu}$ and accordingly the non-trivial center elements read
\be
\label{1477-ab}
\e^{- 4 \pi \vec{\mu} \vec{H}} = z \, ,
\ee
where $H_1 = T_3 \, , \, H_2 = T_8$ are the generators of the Cartan algebra in the fundamental representation,
normalized to $\tr (T_a T_b) = - \frac{1}{2} \delta_{ab}$. Center symmetry of the effective potential (\ref{305-6}) requires
that it remains invariant under the replacement of the background field
\be
\label{1484-ab}
\va \to \va + \frac{4 \pi}{L} \vec{\mu} \, ,
\ee
where $\vec{\mu}$ represents one of the roots (\ref{1462-G19}).

\section{\label{VG}The Coulomb gauge propagators}

Since $\vp_{\sigma = 0} = \vp$ (see Eq.~(\ref{1048-oo})) by Eq.~(\ref{1060-z2}) the zero-temperature propagators
in Coulomb gauge (for vanishing background field) are given by the background gauge propagators
 with $\sigma = 0$. Putting $\sigma = 0$ in Eq.~(\ref{1031-z1}) the zero-temperature gap equation in Coulomb gauge becomes in the presently used approximation
\be
\label{1360-x2}
\omega^2 (\vp) = \vp^2 + \chi^2 (\vp) \,.
\ee
Since $\chi (\vp) = \chi^{\sigma = 0} (\vp)$ (\ref{1206-x4}) is defined in terms of the ghost propagator $G^\sigma (\vp)$ this 
equation has to be solved simultaneously with the ghost DSE (\ref{1183-x1}). Introducing the ghost form factor $d (\vp)$ by
\be
\label{1366-aa}
G^{\sigma = 0} (\vp) = \frac{d (\vp)}{g\, \vp^2}
\ee
the ghost DSE becomes
\begin{gather}
\label{1371-G26}
d^{- 1} (\vp) = \frac{1}{g} + I_d (\vp) \,,\\
 I_d (\vp) = N_c \int \dbar{q} \bigl[ 1 -  (\hat{\vp} \hat{\vq})^2 \bigr]
\frac{d(\vp - \vq)}{(\vp - \vq)^2} \frac{1 }{2 \omega (\vq)}  \nonumber\, .
\end{gather}
Equations (\ref{1360-x2}) and (\ref{1371-G26}) can be solved analytically in the IR using the power law ans\"atze \cite{Feuchter:2004mk,Schleifenbaum:2006bq,R16} 
\be
\label{1376-aa}
\omega (p) = \frac{A}{p^\alpha} \, , \qquad  d (p) = \frac{B}{p^\beta} \, ,
\ee
with $p = \abs{\vp}$. From the ghost DSE (\ref{1371-G26}) one finds the sum rule
\be
\label{1381-aa}
\alpha = 2 \beta- d + 2 \, 
\ee
\begin{figure}
 \includegraphics[width=0.85\linewidth]{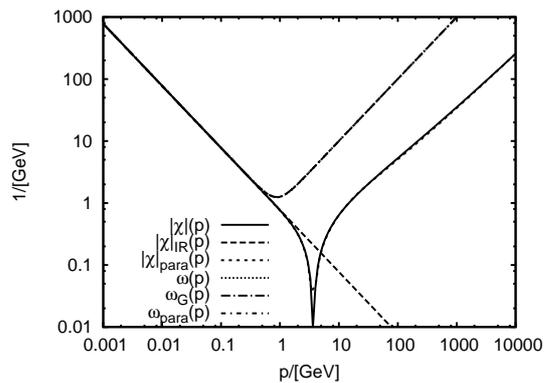}
\caption{The gluon energy $\omega(p)$ and modulus of the curvature $\chi(p)$ resulting from the full numerical solution of
the variational approach in Coulomb gauge as described in Ref.~\cite{R16}.
The Figure contains also the Gribov formula (\ref{1353-x1}) fitted to the numerical data for $\omega(p)$, as well as the IR form $\chi_\mathrm{IR}$ (\ref{1365-x3}) of the curvature. In addition we also show the parameterization (\ref{1423-x1}) $\chi_\text{para}(p)$ to the numerical results for $\chi(p)$, together with the solution 
$\omega_\text{para} (p)$ of the gap equation (\ref{1360-x2}) assuming the optimal parameterization $\chi_\mathrm{para} (p)$ (\ref{1423-x1}) as input.}
\label{fig1}
\end{figure}%
For $d = 3$ the gap equation (\ref{1360-x2}) allows then for two solutions with ghost IR exponents
\be
\label{1386-aa}
\beta = 1, \qquad \beta = 0.795 \, .
\ee
The first one, for which $\alpha = 1$, 
is compatible with the gluon propagator measured on the lattice. Indeed,
the lattice results \cite{R25} for the gluon energy $\omega (p)$ can be well fitted by the Gribov formula \cite{R29}
\be
\label{1353-x1}
\omega_\mathrm{G} (p) = \sqrt{p^2 + \frac{M^4}{p^2}}
\ee
with a mass $M \simeq 880$ MeV for SU($2$) \cite{R25} and about the same value for SU($3$) \cite{RX5}.
Using for $\omega (p)$ the Gribov formula (\ref{1353-x1}) we find from the gap equation (\ref{1360-x2}) for the curvature
\be
\label{1365-x3}
\chi (p) = \frac{M^2}{p} \eqcolon \chi_\mathrm{IR} (p) \, ,
\ee
which represents the correct infrared behavior of $\chi (p)$ obtained in Refs.~\cite{Feuchter:2004mk,Schleifenbaum:2006bq,Reinhardt:2007wh}. What is missing in the expression (\ref{1365-x3}) is the UV-part
\be
\label{1371-x4}
\chi_\mathrm{UV} (p) = c \frac{\sqrt{p^2 + \lambda}}{\ln \frac{p^2 + \lambda}{M^2}} \, \xrightarrow{p \to \infty}\,  c \frac{p}{\ln \frac{p^2}{M^2}} , \quad c = \text{const} \, ,
\ee
which arises in a self-consistent solution of the coupled ghost DSE (\ref{1371-G26}) and gap equation (\ref{1360-x2})
 \cite{Feuchter:2004mk},\cite{R16}. In the gap equation (\ref{1360-x2}) this term is UV-subleading compared to the UV-leading term $\omega (p) \sim p$.

Alternative to the use of the lattice result we solve the ghost DSE (\ref{1371-G26}) and the gap equation
(\ref{1360-x2}) numerically in the way described in Ref.~\cite{R16}, 
where also the 
renormalization of these equations is discussed in detail. 
The resulting numerical solutions for $\omega (p) = \omega^{\sigma = 0} (p)$ and $\chi (p) = \chi^{\sigma = 0} (p)$ 
are shown in Fig.~\ref{fig1} for the $\beta = 1$ solution. The physical scale is determined 
as in Ref.~\cite{R16} by fitting the numerical results for $\omega (p)$ to the Gribov formula 
(\ref{1353-x1}), 
using the lattice result of 
$M = 880$ MeV as input. Fig.~\ref{fig1} shows also the fit of the the Gribov formula (\ref{1353-x1}) to the numerical results for $\omega (p)$ and $\chi_\mathrm{IR} (p)$ (\ref{1365-x3}). In the double logarithmic plot the Gribov formula (\ref{1353-x1}) is indistinguishable from the numerical results.

For practical reason in the evaluation of the effective potential of the background field we parameterize the numerical results
for $\chi (p)$ by
\be
\label{1423-x1}
\chi_\mathrm{para} (p) = u (p) \chi_\mathrm{IR} (p) +  v (p) \chi_\mathrm{UV} (p) \, ,
\ee
where $\chi_\mathrm{IR} (p)$ and $\chi_\mathrm{UV} (p)$ are defined by Eqs.~(\ref{1365-x3}) and (\ref{1371-x4}), and $u (p)$ and $v (p)$ are smooth cut-off functions
to restrict $\chi_\mathrm{IR} (p)$ and $\chi_\mathrm{UV} (p)$ to the IR and UV-regime, respectively. An optimal fit to the numerical results for
$\chi (p)$ is obtained with the choice
\be
\label{1430-asa}
u (p) = \left[ 1 + \lk \frac{p}{\eta} \rk^m \right]^{- 1} \,, \quad 
v (p) = 1 - \lk \frac{\lambda}{p^2 + \lambda} \rk^n \, .
\ee
From the best fit of Eq.~(\ref{1423-x1}) to the numerical data one extracts the following values for the parameters (for SU($2$))
\be
\label{1436-aa}
\begin{aligned}
\eta & = 4.83 M,   & \lambda &= 89.62 M^2 \\
m & = 3.13,   & n &= 0.89 \, .
\end{aligned}
\ee
Fig.~\ref{fig1} shows also the fitted curve (\ref{1423-x1}) together with the numerical data for $\chi (p)$.  
With $\omega (p)$ and $\chi (p)$ at hand we are now in a position to evaluate the effective potential of the Polyakov loop, see Eq.~(\ref{1088-l6}) below. 

\section{\label{sectVI}The effective potential of the Polyakov loop}

As discussed in Sect.~\ref{sectionII-neu} a constant background field $a$ living in the Cartan algebra can serve as order parameter
for confinement when it is directed along a compactified dimension. Thereby the inverse of the length $L$ of the compactified
dimension figures as temperature. In the Hamiltonian approach the effective potential of a spatial background field $a$ is given by the energy 
density \cite{WeinbV2}
\be
\label{1359-G236}
e (a, L) = \frac{\vev{ H }_a}{V (d - 1)} \, ,
\ee
where $\vev { H }_a$ is defined by the constrained variational problem (\ref{603-G9}) and given in the present approach by Eq.~(\ref{1212-18}). Furthermore, $V$ is the spatial volume and for convenience we have divided the energy by the number of 
transversal spatial dimensions.

Assuming $d = 3$ and compactifying the $3$-axis
to a circle with circumference $L$ the above formulae all remain valid except for 
the following modifications: The integration over coordinate space reads
\be
\label{1056-l1}
\int \d[3] x f (\vx) = \int \d[2] x_\perp \int^L_0 \d x_3 f (\vx_\perp, x_3) \, ,
\ee
where $\vx_\perp$ denotes the projection of $\vx$ onto the 1-2-plane.
Accordingly the integration over momentum space becomes
\be
\label{1062-l2}
\int \dbar[3] p f (\vp) = \int \dbar[2] p_\perp \frac{1}{L} \sum^\infty_{n = - \infty} f 
(\vp_\perp, p_n) \, ,
\ee
where 
\be
\label{1068-l3}
p_n = \frac{2 \pi n}{L}
\ee
are the Matsubara frequencies and the shifted momentum variable $\vp_\sigma$ (\ref{1344-G3}) reads 
\be
\label{1073-l4}
\vp_\sigma = \vp_\perp + \lk p_n - \vec{\sigma} \va \rk \ve_3 \,, 
\ee
With Eq.~(\ref{1062-l2}) we find from Eq.~(\ref{1212-18}) for the energy density (\ref{1359-G236}),
using (\ref{1026-oo}), (\ref{1043-oo}) and (\ref{1054*oo})
\bea
\label{1088-l6}
e (a, L) & =& \sum_\sigma \int \dbar[3] p \lk \omega^\sigma (\vp) - \chi^\sigma (\vp) \rk \nonumber\\
& = & \sum_\sigma \int \dbar[2] p_\perp \frac{1}{L} \sum_n \lk \omega (p_\sigma) - \chi (p_\sigma) \rk \, ,
\eea
where $p_\sigma = \abs{\vp_\sigma}$ is defined by Eq.~(\ref{1073-l4}). By shifting the summation index $n$ one easily verifies
that by Eq.~(\ref{1368-20}) the potential (\ref{1088-l6}) indeed has the periodicity property
\be
\label{1095-l7}
e \lk \va + 4 \pi/L \vmu, L \rk = e (\va, L) 
\ee
required for the potential of the Polyakov loop by center symmetry.

For later use we calculate first the contributions to the effective potential 
arising from a gluon energy $\omega (p_\sigma)$ that obeys a 
power law
\be
\label{1128-l10}
\omega_\alpha (p, \lambda) = M^{1 - \alpha} \lk \sqrt{p^2 + \lambda} \rk^\alpha \, .
\ee
For later convenience we have introduced here an addition parameter $\lambda \geq 0$, which of course is irrelevant for 
$p_\sigma \to \infty$. Furthermore,
 $M$ is a constant of dimension mass, which was introduced
for dimensional reasons. For the $\beta = 1$ solution of the variational calculation (see Sect.~\ref{VG}) the IR and UV behavior of $\omega (k)$ is given by 
$\omega_{\alpha = - 1} (p, \lambda = 0)$ and $\omega_{\alpha = 1} (p, \lambda = 0)$, respectively.
\begin{figure*}[t]
 \subfigure[]{\label{fig2-1}\includegraphics[width=0.4\linewidth]{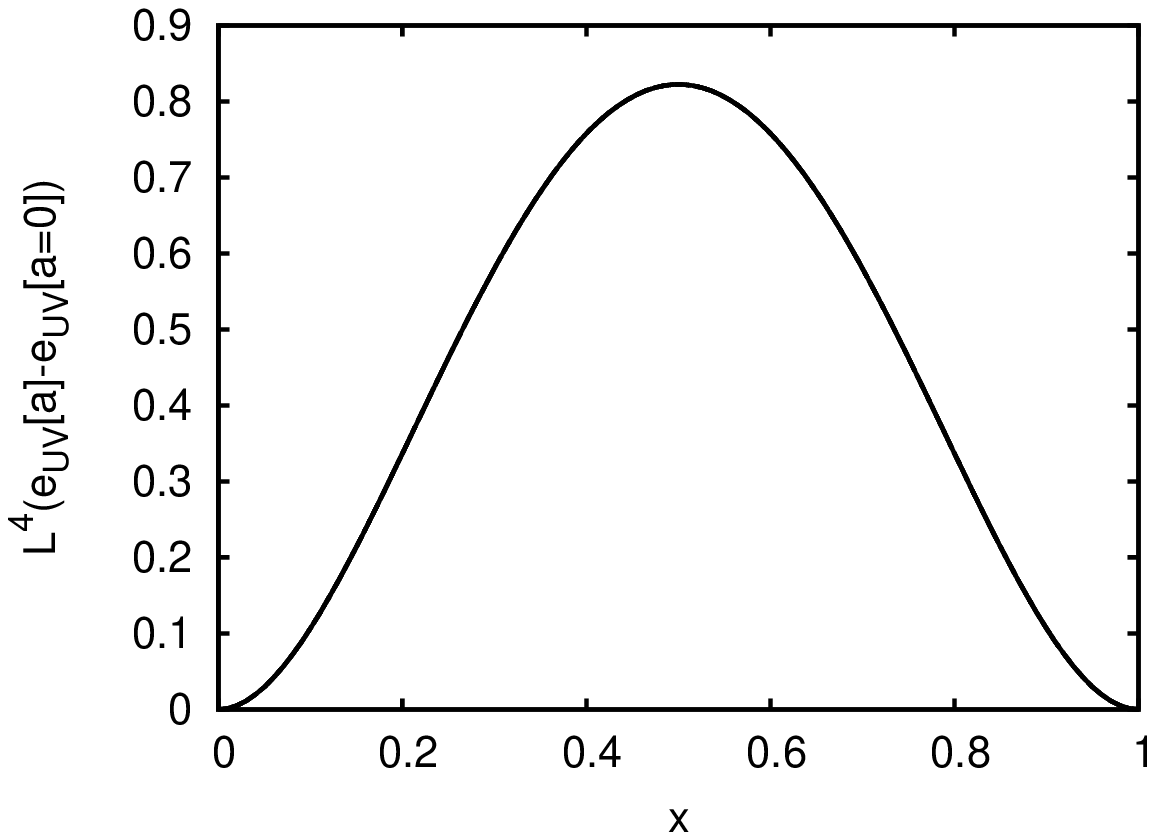}}
 \subfigure[]{\label{fig2-4}\includegraphics[width=0.4\linewidth]{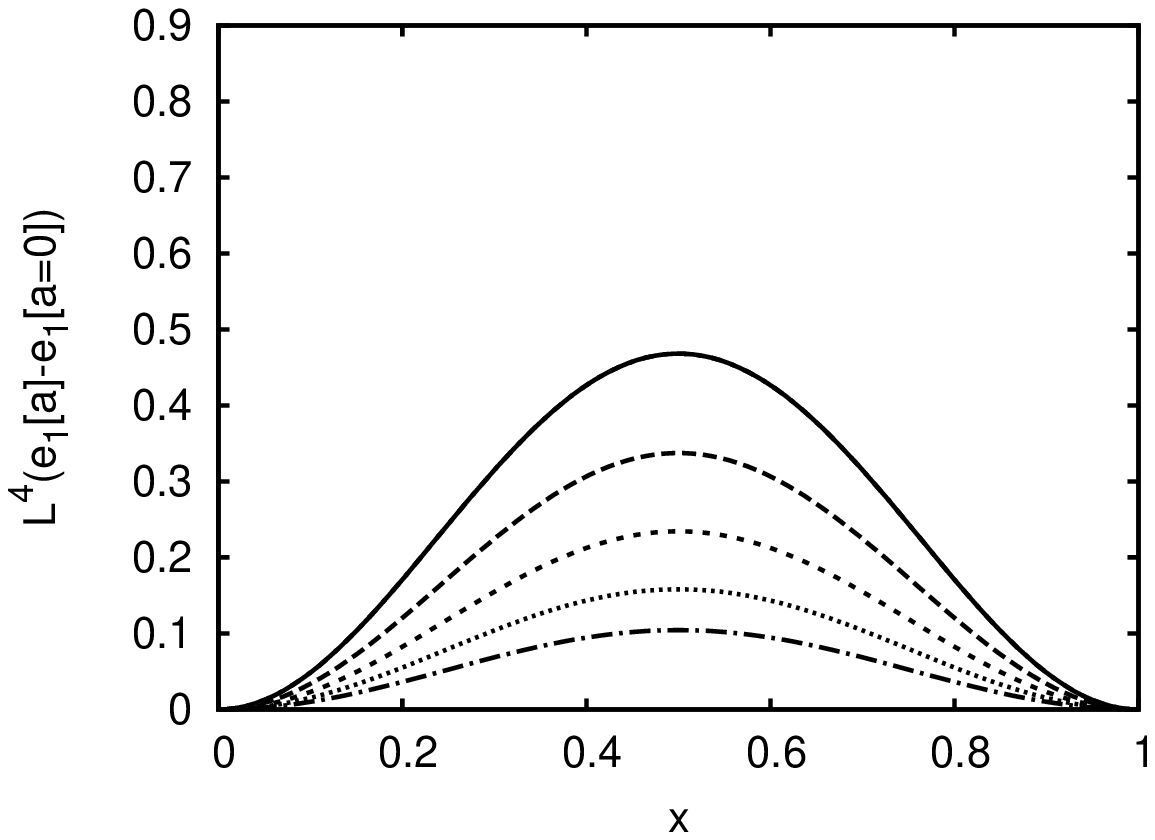}}
 \subfigure[]{\label{fig2-2}\includegraphics[width=0.4\linewidth]{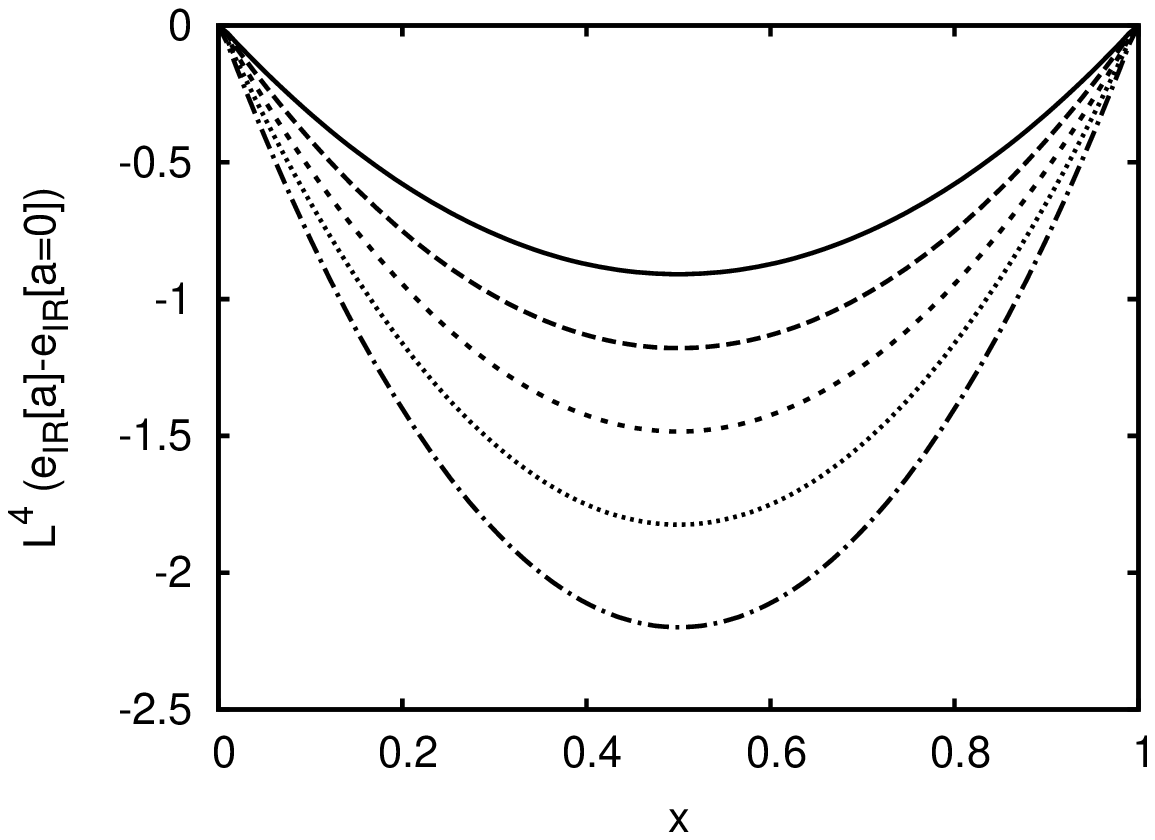}}
 \subfigure[]{\label{fig2-3}\includegraphics[width=0.4\linewidth]{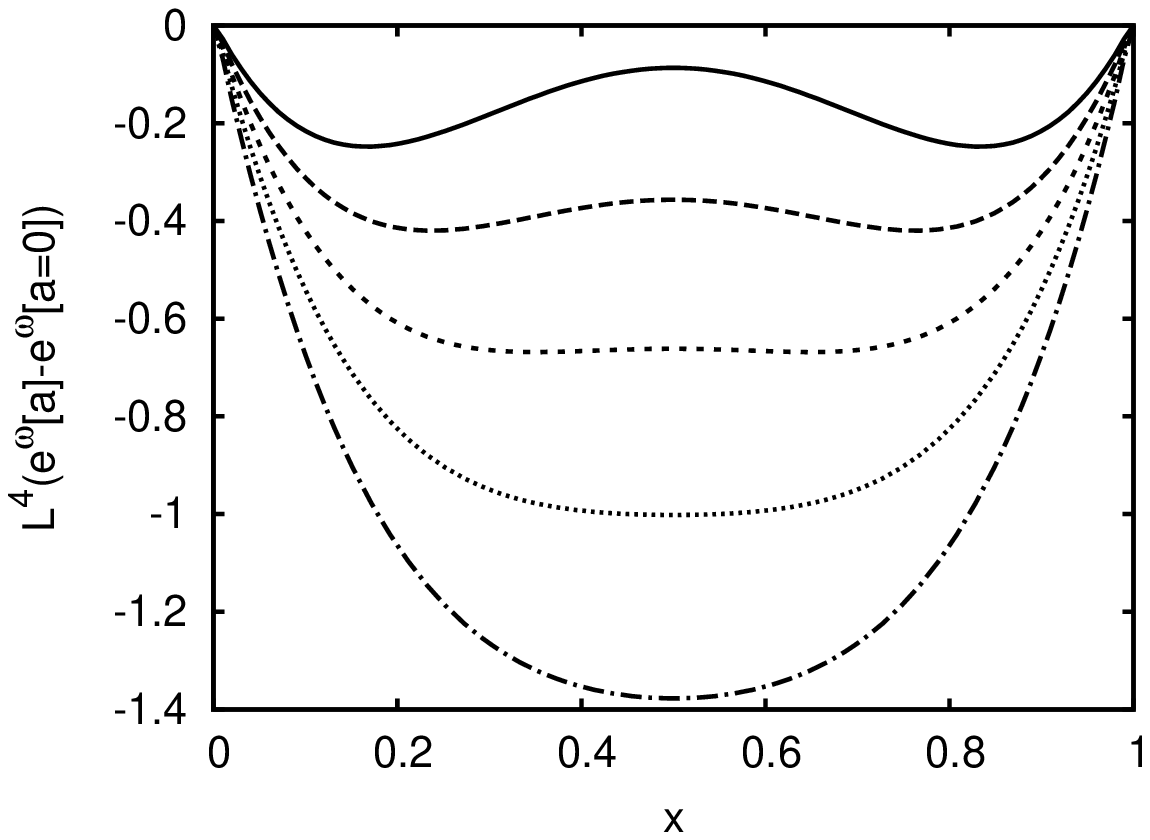}}
\caption{The effective potential $a(a,L)$ (\ref{1088-l6}) multiplied by $L^4$ at different temperatures $L^{- 1}$ for the gauge group SU($2$). The potential is
shown as function of the dimensionless variable $x$ (\ref{1775-29}), where $\va \vsigma = a_3 = a$ for SU($2$), and for various forms of the 
gluon energy $\omega (\vp)$ and neglecting the ghost loop, $\chi (\vp) = 0$: (a) UV-(Weiss) potential (\ref{1686-G31}) $\omega (p) = p$, (b) 
massive dispersion relation (\ref{1772-ii-4-14}) $\omega (p) = \sqrt{m^2 + p^2}$, (c) IR-potential (\ref{1715-aa}) $\omega (p) = M^2 / p$, 
(d) the sum of the UV and IR potential, $\omega (p) = p + M^2 / p$, which is obtained by adding the two potentials shown in (a) and (c).}
\end{figure*} 

The contribution of $\omega_\alpha (p_\sigma, \lambda)$ to the energy density (\ref{1088-l6})
\be
\label{1118-l9}
e_\alpha (a, L, \lambda) \coloneq \sum_\sigma \int \dbar[2] p_\perp \frac{1}{L} \sum_n \omega_\alpha (p_\sigma, \lambda)
\ee
is calculated in appendix \ref{appB}. One finds after the value of the potential at vanishing background field is subtracted, see Eq.~(\ref{1837-ii-13-alt}),
\begin{multline}
\label{1837-ii-13}
\bar{e}_\alpha  (a, L, \lambda ) \coloneq
e_{\alpha} (a, L, \lambda ) - e_{\alpha} (0, L, \lambda)  \\
= -  \frac{8M^{1 - \alpha}}{(4 \pi)^{3/2} \Gamma \lk- \frac{\alpha}{2} \rk} \\
\times \sum_\sigma \sum^\infty_{n = 1}  \lk \frac{2  \sqrt{\lambda}}{n L}\rk^{\frac{\alpha}{2}+\frac{3}{2}} \sin^2 
\left(\frac{ n \vec{\sigma} \va L }{2} \right) K_{-\frac{\alpha}{2}-\frac{3}{2}}(n L \sqrt{\lambda})\, .
\end{multline}
where $K_r (z)$ is the modified Bessel function (\ref{1666-ac}). For $\lambda = 0$ the above expression simplifies to
(see Eq.~(\ref{1837-ii-12-alt}))
\begin{multline}
\label{1837-ii-12}
\bar{e}_{\alpha} (a, L, \lambda = 0)  = -  \frac{8 M^{1 - \alpha}}{(4 \pi)^{3/2}} \frac{\Gamma \lk 
\frac{3}{2} + \frac{\alpha}{2} \rk}{\Gamma \lk
- \frac{\alpha}{2} \rk}  \\
\times \lk \frac{2}{L} \rk^{3 + \alpha} \sum_{\vec{\sigma} > 0} 
\sum^\infty_{n = 1} \frac{1}{n^{3 + \alpha}} \sin^2 \left(\frac{ n \vec{\sigma} \va L }{2}\right) \, ,
\end{multline}
where the summation is now over the positive roots only.

Before we give a full numerical calculation of the effective potential $e (a, L)$ (\ref{1088-l6}) let us discuss its
qualitative features to reveal how the deconfinement phase transition shows up in this quantity. For this
purpose we calculate below the IR and UV contributions to the effective potential. Note also that the momentum integrals
(\ref{1088-l6}) are UV divergent, so that the UV contributions require an analytic treatment to remove the divergences.

\subsection{Asymptotic contributions}

For a first 
qualitative discussion of the Polyakov loop potential let us ignore the curvature $\chi (p)$ in Eq.~(\ref{1088-l6}). 
$\chi (p)$ is UV-subleading but has the same IR behavior as $\omega (p)$, see Sect.~\ref{VG}.
Therefore, we expect a substantial change of the deconfinement phase transition temperature by ignoring
$\chi (p)$. Nevertheless, in order to see how the deconfinement phase transition manifests itself in the
effective potential it is enlightening to consider the simplified potential arising from 
(\ref{1088-l6}) when the curvature $\chi (p)$ is ignored. We first calculate the effective potential arising from
the UV behavior of $\omega (p)$.

In the ultraviolet the gluon energy is given by $\omega (p) = |\vp|$, which follows from Eq.~(\ref{1128-l10}) with $\alpha = 1$ and $\lambda = 0$. For 
these values we obtain
from Eqs.~(\ref{1118-l9}), (\ref{1837-ii-12})
\begin{multline}
\label{1211-21}
\bar{e}^\omega_\mathrm{UV} (a, L) \coloneq e_{\alpha = 1} (a, L, \lambda = 0) - e_{\alpha = 1} (0, L, \lambda = 0)\\
 = \frac{8}{\pi^2} \frac{1}{L^4} \sum_{\sigma > 0} \sum^\infty_{n = 1}
\frac{\sin^2 \lk  n \va \vec{\sigma} {L}/{2} \rk}{n^4} \, .
\end{multline}
The summation over the Matsubara frequencies can be carried out for $0 \leq \va \vsigma L / 2 \pi \leq 1$ using
\be
\label{1681-aa}
\sum^\infty_{n = 1} \frac{\cos n x}{n^4} = \frac{\pi^4}{90} - \frac{\pi^2 x^2}{12} + \frac{\pi x^3}{12} - \frac{x^4}{48} \, , 
\quad 0 \leq x \leq 2 \pi
\ee
which yields
\be
\label{1686-G31}
\bar{e}^\omega_\mathrm{UV} (a, L) = \frac{4}{3} \frac{\pi^2}{L^4} \sum_{\sigma > 0} \lk \frac{\va \vec{\sigma}  L}{2 \pi} 
\rk^2 \left[ \frac{\va  \vec{\sigma} L}{2 \pi} - 1 \right]^2 \, .
\ee
This is precisely the Weiss potential, which is shown in Fig.~\ref{fig2-1} for the gauge group SU($2$). It was originally 
obtained in Ref.~\cite{Weiss:1980rj} in a 1-loop calculation in Landau gauge. Thus, the
present quasi-particle type approximation (\ref{1118-l9})  to the gluons (cf. also Eq.~(\ref{ex-11}))
with the gluon energy $\omega (p)$ replaced by its ultraviolet limit, i.e. by the photon energy $\omega_{\alpha = 1} (p, \lambda = 0) = \abs{\vp}$,
is equivalent to an ordinary $1$-loop background calculation. Since up to the periodicity (\ref{305-6}) the Weiss potential is minimal only for $a = 0$ we find for the
Polyakov loop
\be
\label{1735-ad}
\langle P [A_3] \rangle \simeq P \left[ \langle A_3 \rangle = a = 0 \right] = 1
\ee
indicating the deconfined phase.
\begin{figure}
 \subfigure[]{\label{figN3-1a}\includegraphics[width=0.85\linewidth]{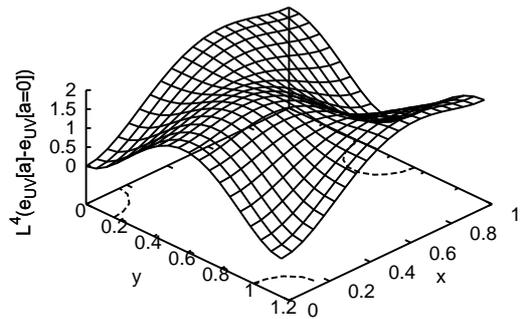}}
 \subfigure[]{\label{figN3-1b}\includegraphics[width=0.85\linewidth]{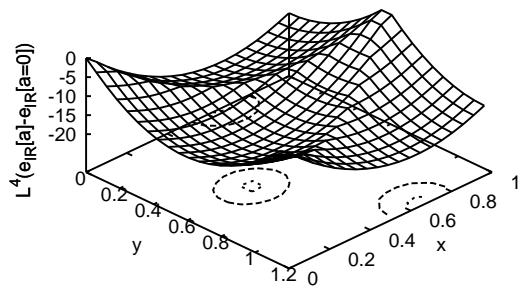}}
\caption{(a) The Weiss potential (\ref{1686-G31}) and (b) the infrared potential (\ref{1715-aa}) at  $L^{-1} = 290$ MeV for the gauge group SU($3$)
 as a function of the dimensionless fields variables (\ref{1775-29}) $x$, $y$ and multiplied by $L^4$.  In both plots the local minima are marked by 
 contours in the $xy$-plane.}
\label{figN3-1}
\end{figure}

For the sake of illustration we also consider a massive gluon dispersion relation
\be
\label{massgrib}
  \omega(p) =\sqrt{M^2 + p^2}\,,
\ee
which is obtained in the variational approach to Yang--Mills theory in Coulomb gauge above the critical temperature 
of the deconfinement phase transition \cite{R16}. This dispersion relation follows from Eq.~(\ref{1128-l10}) for 
$\alpha =1$ and $\lambda = M^2$. With these  parameter values we find from Eq.~(\ref{1837-ii-13}) for the effective 
potential 
\begin{multline}
\label{1772-ii-4-14}
\bar{e}_1 (a, L, M^2)\coloneq e_1 (a, L, M^2)-e_1 (0, L, M^2) \\
= \frac{M^2}{2 \pi^2} \sum_\sigma \sum^\infty_{n=1} \left(\frac{2}{n L} \right)^2 \sin^2\left(n \vec{\sigma} \va L/2 \right) K_{-2}(n L M)\,.
\end{multline}
This potential is show in Fig.~\ref{fig2-4} for SU($2$). For $M=0$ this potential reduces, of course, to the Weiss potential 
shown in Fig.~\ref{fig2-1}. Due to the presence of the energy scale $M$ in Eq.~(\ref{massgrib}) this potential is now temperature dependent. (The Weiss potential $L^4 \bar{e}^\omega_\text{UV}$ is independent of $L$, see Fig.~\ref{fig2-1}.)
For any value of $M$ this potential is minimal at $a=0$ and gives rise to the same Polyakov loop as the Weiss potential.

The quasi-particle approximation to the energy density is, however, a priori non-perturbative. The gluon 
energy $\omega (\vp)$ determined in the variational approach, in particular, captures the infrared confinement
properties of the gluons. In the infrared the gluon energy (\ref{1353-x1}) is given by Eq.~(\ref{1128-l10}) with $\alpha = - 1$  and $\lambda = 0$.
For these values we find from Eq.~(\ref{1837-ii-12})
\begin{multline}
\label{1219-22}
\bar{e}^\omega_\mathrm{IR} (a, L) \coloneq e_{\alpha = - 1} (a, L, \lambda = 0) - e_{\alpha = - 1}  (0, L, 
\lambda = 0)\\
 = - 4 \frac{M^2}{\pi^2} \frac{1}{L^2} \sum_{\sigma > 0} \sum^\infty_{n = 1} \frac{1}{n^2} \sin^2 \lk
n \va \vec{\sigma} {L}/{2} \rk \, .
\end{multline}
The summation over $n$ can be carried out for $0 \leq \va \vsigma L / 2 \pi < 1$ using
\be
\label{1710-aa}
\sum^\infty_{n = 1} \frac{\cos (n x)}{n^2} = \frac{\pi}{6} - \frac{\pi x}{2} + \frac{x^4}{4} \, , \quad 0 \leq x \leq 2 \pi \, ,
\ee
which yields
\be
\label{1715-aa}
\bar{e}^\omega_\mathrm{IR} (a, L) = 2 \frac{M^2}{L^2} \sum_{\sigma > 0} \left[ \lk \frac{\va \vec{\sigma}L}{2 \pi} \rk^2 - \frac{\va
\vec{\sigma} L}{2 \pi} \right] \, .
\ee
This potential is shown in Fig.~\ref{fig2-2} for the gauge group SU($2$). It drastically differs from the Weiss potential (\ref{1211-21}) obtained from the UV behavior of $\omega (p)$ and shown in Fig.~\ref{fig2-1}.
First it has the opposite sign and second a non-trivial $L$-dependence. Its minimum occurs at $a = \frac{\pi}{L}$
corresponding to a center symmetric ground state. Accordingly it yields a vanishing expectation value of the
Polyakov loop (\ref{1735-ad}).
\begin{align}
\label{1780-ad}
\langle P [A_3] \rangle &\simeq P\left[\langle A_3 \rangle =a = \frac{\pi}{L}\right] = 0\, .
\end{align}
In Figs.~\ref{figN3-1} (a) and (b) the asymptotic UV and IR potentials (\ref{1686-G31}) and (\ref{1715-aa}), respectively,
 are shown for the gauge group SU($3$) as function of the dimensionless variables
\be
\label{1775-29}
x \coloneq \frac {a_3 L}{2 \pi}\,, \quad y \coloneq \frac {a_8 L}{2 \pi}\,.
\ee
By the periodicity of the effective potential these variables can be restricted to the intervals
\be
   0 \leq x \leq 1\,, \quad  0 \leq y \leq \frac{2}{\sqrt{3}}\,.
\ee
As mentioned before, the Weiss potential $L^4 \bar{e}^\omega_\text{UV}(a,L)$ is independent of the temperature $L^{-1}$, while the IR potential is temperature dependent and is shown in Fig.~\ref{figN3-1b} for $L^{-1}= 290$ MeV. The minima of the Weiss potential are degenerate and occur at field configurations $a_\text{min}$, for which the Polyakov line yields a center element $z$
\be
\e^{-L a_\text{min}} = z\; \in \; \mathrm{Z}(N)\,,
\ee
and thus $P[a_\text{min}] = z \neq 0$ as expected for the deconfined phase.
For SU($3$) the minima of the Weiss potential occur at (see Fig.~\ref{figN3-1a})
\be
 \lk x_\text{min},\; y_\text{min} \rk = \lk 0,\, 0\rk, \;\lk 0,\, \frac{2}{\sqrt{3}}\rk, \; \lk 1,\,  \frac{1}{\sqrt{3}}\rk\,
\ee
for which the Polyakov line is
\be
\e^{-L a_\text{min}} = \Id,\quad  \e^{\i \frac{2 \pi }{3}} \Id,\quad  \e^{-\i \frac{2 \pi}{3}} \Id\,.
\ee
The minima of the confining IR potential occur at center symmetric configurations, for which the Polyakov loop vanishes. For SU($2$) this configuration is given by $x_\text{min} =1/2$ (see Fig.~\ref{fig2-2}) while for SU($3$) these configurations read (see Fig.~\ref{figN3-1b}):
\be
  \lk x_\text{min}, y_\text{min} \rk = \lk \frac{2}{3},\, 0\rk, \;\lk \frac{2}{3},\, \frac{2}{\sqrt{3}}\rk, \; \lk \frac{1}{3},\,  \frac{1}{\sqrt{3}}\rk\,
\ee
and the corresponding Polyakov lines are given by
\begin{align}
\e^{-L a_\text{min}} = &  \text{diag} \lk \e^{\i \frac{2 \pi }{3}} ,\;\e^{-\i \frac{2 \pi }{3}},\;1 \rk \,, \nonumber \\
&\text{diag} \lk \e^{-\i \frac{2 \pi }{3}} ,\;1 ,\; \e^{\i \frac{2 \pi }{3}}\rk \, ,\nonumber \\
& \text{diag} \lk \e^{\i \frac{2 \pi }{3}} ,\;1 ,\; \e^{-\i \frac{2 \pi }{3}}\rk 
\end{align}
which all yield $P[a_\text{min}] =0$.
\subsection{The deconfinement phase transition}

Clearly the deconfinement phase transition is
related to the transition in the effective potential from its IR behavior (\ref{1219-22}) to its UV behavior (\ref{1211-21}). 
To illustrate this let us approximate 
the
Gribov formula  (\ref{1353-x1})
by
\begin{figure}
 \includegraphics[width=0.85\linewidth]{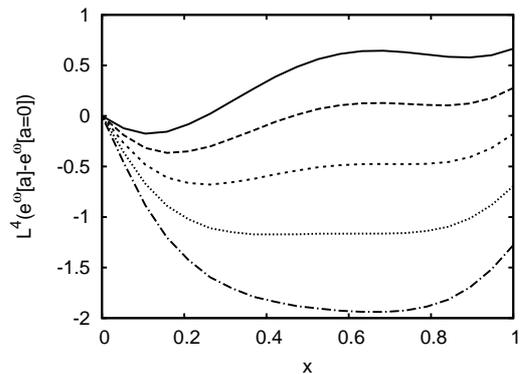}
 \caption{The $a_8 = 0$ slice of the potential (\ref{1241-25}) for the gauge group SU($3$)  as a function of $x$ (\ref{1775-29}) for different $L^{-1}$ in the range from $484$ MeV to $880$ MeV (from bottom to top). The critical temperature is $597$ MeV.}
\label{figN3-2}
\end{figure}
\be
\label{1236-24}
\sqrt{p^2 + \frac{M^4}{p^2}} \longrightarrow p + \frac{M^2}{p}  \, ,
\ee
i.e. we choose the gluon energy as
\be
\label{1228-23}
\omega (p) = \omega_{\alpha = 1} (p, \lambda = 0) + \omega_{\alpha = - 1} (p, \lambda = 0) \, .
\ee
This expression is correct in both the IR and UV but certainly introduces an error in the mid-momentum regime, which influences the 
deconfinement phase transition. 
With Eq.~(\ref{1228-23}) the energy density (\ref{1088-l6}) for $\chi (p) = 0$ becomes 
\begin{multline}
\label{1241-25}
\bar{e}^\omega (a, L)  =  \bar{e}^\omega_\text{IR} (a, L) + \bar{e}^\omega_\text{UV} (a, L) \\
 = \frac{4}{\pi^2} \frac{1}{L^4} \sum^\infty_{n = 1} \frac{1}{n^2} \left[ \frac{2}{n^2} - (M L)^2 \right] \sum_{\sigma > 0}  \sin^2 \lk n 
{\va \vec{\sigma}  L}/{2} \rk
\end{multline}
For SU($2$) the only positive root is $\sigma =1$ and this potential can be expressed as
\be
 \bar{e}^\omega(a,L)=  \frac{4}{3} \frac{\pi^2}{L^4} f (x) , \quad f (x) = x^2 (x - 1)^2 + c (x^2 - x)   \, ,
\ee
where
\be
\label{1745-aa}
c \coloneq \frac{3 M^2 L^2}{2 \pi^2}
\ee
and the variable $x$ is defined in Eq.~(\ref{1775-29}). For small temperatures $L^{- 1} \ll M$ this potential is negative and the system is in the confined phase. As $L^{- 1}$ 
increases the  minimum of the potential at $x = \frac{1}{2}$ turns into a maximum
 and the system makes a transition to the deconfined phase, see Fig.~\ref{fig2-3}. 
In the relevant interval $0 \leq x \leq 1$ the potential has a single minimum in the confined phase and two degenerate minima in the deconfined phase, which merge to the single minimum at $x = \frac{1}{2}$ of the confined phase at the phase transition. Starting from the deconfined phase
the phase transition occurs where the three roots of 
$f' (x) = 0$ degenerate. This occurs for $c = \frac{1}{2}$, i.e. for
\be
\label{1752-aa}
T_c = L^{- 1} = \sqrt{3} \frac{M}{\pi} \, .
\ee
With the lattice result $M = 880$ MeV this corresponds to $T_c \approx 485$ MeV. 

To exibit the deconfinement phase transition for the gauge group SU($3$) we cut the potential along a line which contains one of the degenerate minima both in the confined and deconfined phase. A convenient cut\footnote{Alternative possible cuts are $y=\frac{1}{\sqrt{3}}$ and $y =\frac{2}{\sqrt{3}}$.} is $y=0$.
Fig.~\ref{figN3-2} shows the $y = 0$ (i.e. $a_8 = 0$) cut through the potential (\ref{1241-25}). At low temperatures the minimum corresponds to the center symmetric configuration $x = \frac{2}{3}$ ($y = 0$). At a certain critical temperature this minimum disappears while a new minimum at a smaller $x$-value  occurs, which (for sufficiently high temperature) eventually becomes the minimum $x = y = 0$ for which the Polyakov loop equals the trivial center element $z = \Id$.
From Fig.~\ref{figN3-2} we find a critical temperature of $T_c \approx 597$ MeV.

The critical tempertures obtained above are by far too high. Of course, given the approximations used to arrive at these values
we do not expect a decent description of $T_c$. These approximations 
are: i) neglect of the curvature $\chi (p)$ and ii) approximating Gribov's formula (\ref{1236-24}). 
The use of the correct Gribov formula (\ref{1353-x1}) only slightly reduces $T_c$ as long as the curvature is neglected as we will show now explicitly for the gauge group SU($2$).

\begin{figure}
\centering
\includegraphics[width=0.85\linewidth]{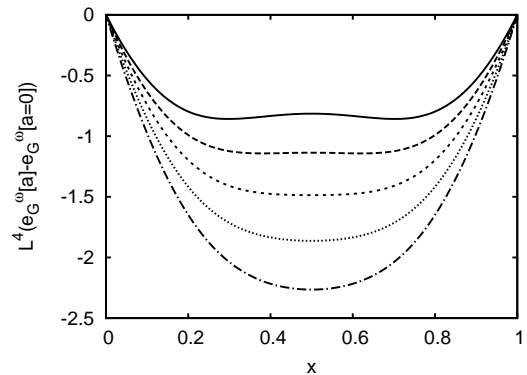}
\caption{The energy density (\ref{1272-28}) for the gauge group SU($2$) as a function of 
the variable $x$ (\ref{1745-aa}) for different $L^{-1}$ in the range from $360$ to $480$ MeV (from bottom to top).  
The phase transition occurs at a temperature of $T_c \approx 432$ MeV.}
\label{fig3}
\end{figure}

Using the Gribov formula (\ref{1353-x1}) for $\omega(k)$ but still neglect the curvature the effective potential (\ref{1088-l6}) reads:
\be
\label{1782-32}
e^\omega_\text{G} (a, L) \coloneq \sum_\sigma \int \dbar[2] p_\perp \frac{1}{L} \sum_{n=-\infty}^\infty\omega_\text{G} (p_\sigma) \, .
\ee
The evaluation of $e^\omega_\text{G} (a, L)$ has
to be done numerically. To 
avoid UV divergences we define
\be
\bar{e}^\omega_{\mathrm{G}}(a, L)  \coloneq e^\omega_{\mathrm{G}} (a, L)  -e^\omega_{\mathrm{G}} (a=0, L) 
\ee
and represent this quantity as 
\begin{multline}
\label{1272-28}
\bar{e}^\omega_{\mathrm{G}} (a, L)  = \left[ \bar{e}^\omega_\mathrm{G} (a, L) - \bar{e}^\omega_\mathrm{UV} (a, L) \right] + \bar{e}^\omega_\mathrm{UV} (a, L)\\
 =  \sum_\sigma \int \dbar[3]  p \, \Big\{ \left[ \omega_\mathrm{G} (\vp_\sigma) - \omega_{\alpha = 1} (\vp_\sigma, \lambda = 0) \right] \\
 - \left[ \omega_\mathrm{G}(\vp_{\sigma=0}) -\omega_{\alpha =-1}(\vp_{\sigma=0},\lambda=0)  \right]\Big\}+ \bar{e}^\omega_\mathrm{UV} (a, L)  \, 
\end{multline}
and calculate the integral numerically while using the analytic expression (\ref{1686-G31}) in the last term.
The resulting effective potential is shown in Fig.~\ref{fig3} for the gauge group SU($2$) for several temperatures $L^{- 1}$. One finds a critical
temperature for the deconfinement phase transition of $T_c \approx 432$ MeV, which is still much too high.

It is the omission of the
curvature, which pushes the deconfinement phase transition to higher temperatures. This can be seen as follows: In the UV the 
curvature $\chi (p)$ is suppressed by $1/\ln p$ compared to the gluon energy $\omega (p)$ and is in addition negative. Thus
neglecting the curvature at most decreases the UV part of the potential, which is deconfining. In the deep IR the curvature
$\chi (p)$ agrees with $\omega (p)$ (cf. Eqs.~(\ref{1353-x1}) and (\ref{1365-x3})). Neglecting here $\chi ({p})$ definitely increases the contribution of the confining IR
potential. So in total, the neglect of the curvature $\chi (p)$ decreases the deconfining part and at the same time increases
the confining part of the potential. Both effects increase the transition temperature. This will be confirmed by the numerical
results given in the next subsection.
As we will explicitly see below the 
critical temperatures decreases substantially when
the curvature $\chi (p)$ is fully included.

\subsection{The full potential}\label{sect.VIC}
\begin{figure}
 \subfigure[]{\label{fig4}\includegraphics[width=0.85\linewidth]{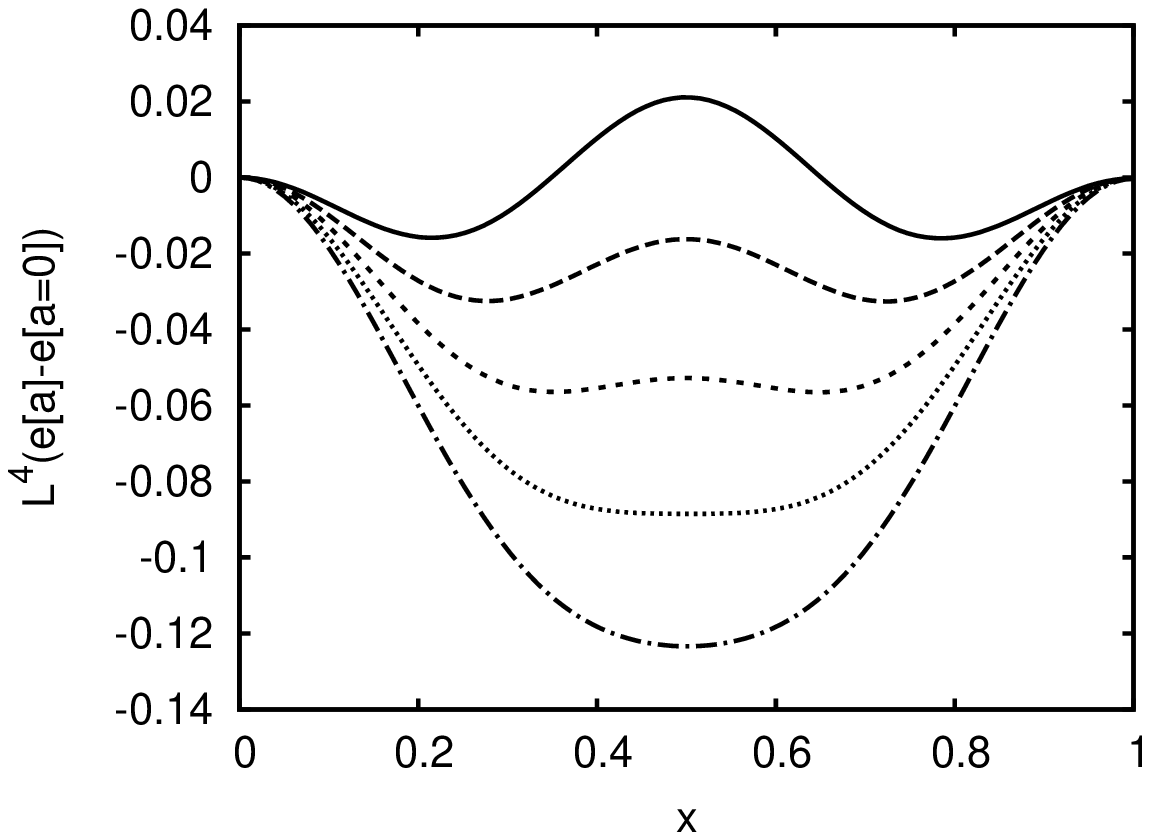}}
 \subfigure[]{\label{figN3-3}\includegraphics[width=0.85\linewidth]{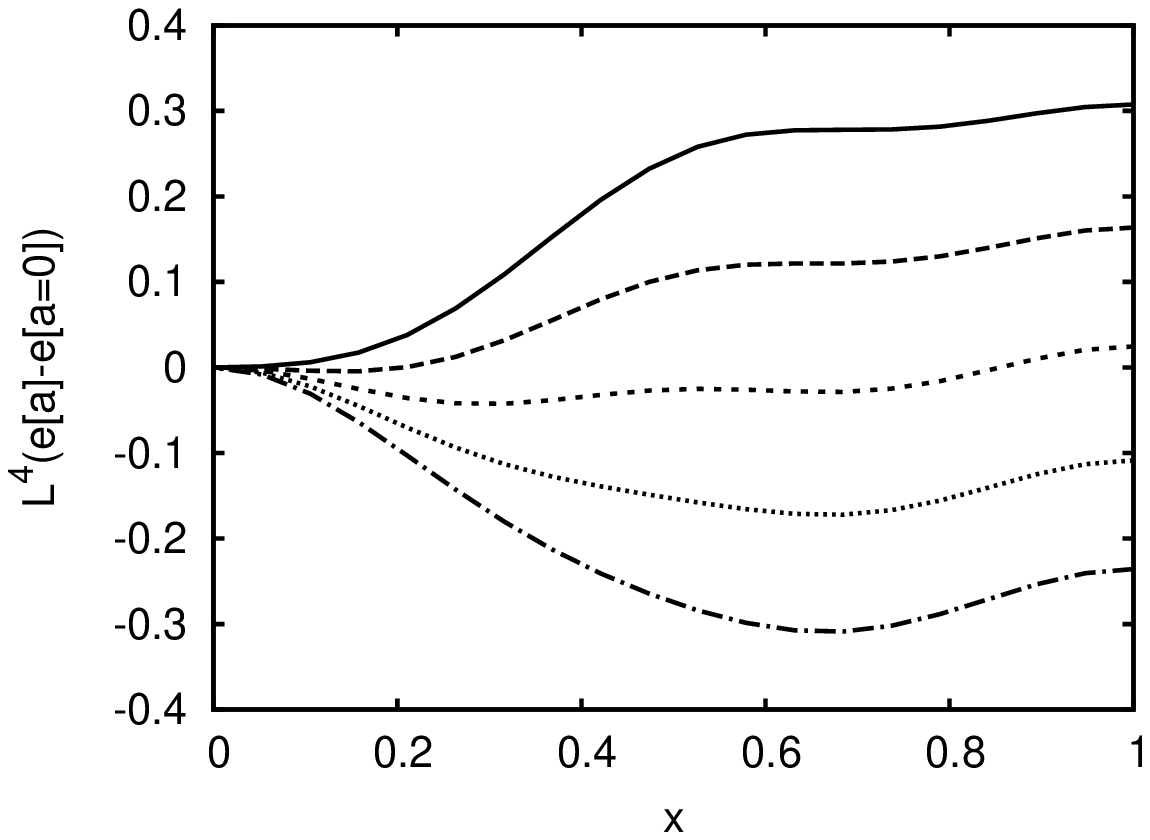}}
\caption{The effective potential (\ref{1803-aa}) for gauge group SU($2$) (a) and the $y=0$ ($a_8 = 0$) slice for SU($3$) (b) as a function of $x$ for different $L^{-1}$ in the range from $260$ to $290$ MeV in figure (a) and from  $260$ MeV to $300$ MeV in figure (b) (from bottom to top).  We extract the following phase transition temperatures: $T_c \approx 267$ MeV (SU($2$)) and $278$ MeV (SU($3$)).}
\label{fig4both}
\end{figure}
Finally, we calculate the full effective potential (\ref{1088-l6}) with the curvature $\chi$ 
included.
We first use the Gribov formula (\ref{1353-x1}) for $\omega (p)$ 
and in accord with the gap equation (\ref{1360-x2}) for $\chi (p)$ the IR expression
$\chi_\text{IR} (p)$ (\ref{1365-x3}), which agrees with the IR behavior of $\omega (p)$. The energy density (\ref{1088-l6}) then becomes
\be
\label{1803-aa}
e (a, L)  = \sum_\sigma \int \dbar[3] p \left[  \omega_\text{G} (\vp_\sigma) - \chi_\text{IR} (\vp_\sigma) \right]  \,.
\ee
Substracting the value at $a=0$ we have
\begin{align}
\bar{e} (a, L) &=  e (a, L)  - e (a, L=0) \nonumber \\
 &= \bar{e}^\omega_\text{G} (a, L) - \bar{e}^\omega_\text{IR} (a, L) \, ,
\end{align}
where $\bar{e}^\omega_\text{G} (a, L)$ and $\bar{e}^\omega_\text{IR} (a, L)$ are defined in Eqs.~(\ref{1272-28}) and (\ref{1219-22}), respectively. (The reader is 
advised to compare this expression to Eq.~(\ref{1241-25})!) The resulting
potentials is shown in Fig.~\ref{fig4} for the gauge group SU($2$), while Fig.~\ref{figN3-3} shows the $a_8 =0$ cut through the SU($3$) potential for various temperatures $L^{- 1}$. 
In both cases one observes a (deconfinement) phase transition, which, however, is first order for SU($2$) and second order for SU($3$). This is because at the deconfinement transition the position of the minimum of the potential changes continuously for SU($2$) but discontinuously for SU($3$).
From these potentials one extracts a critical temperature
of $T_c \approx 267$ MeV and $T_c \approx 277$ MeV for SU($2$) and SU($3$), respectively.
\begin{figure}
\centering
 \subfigure[]{\label{fig7a}\includegraphics[width=0.85\linewidth]{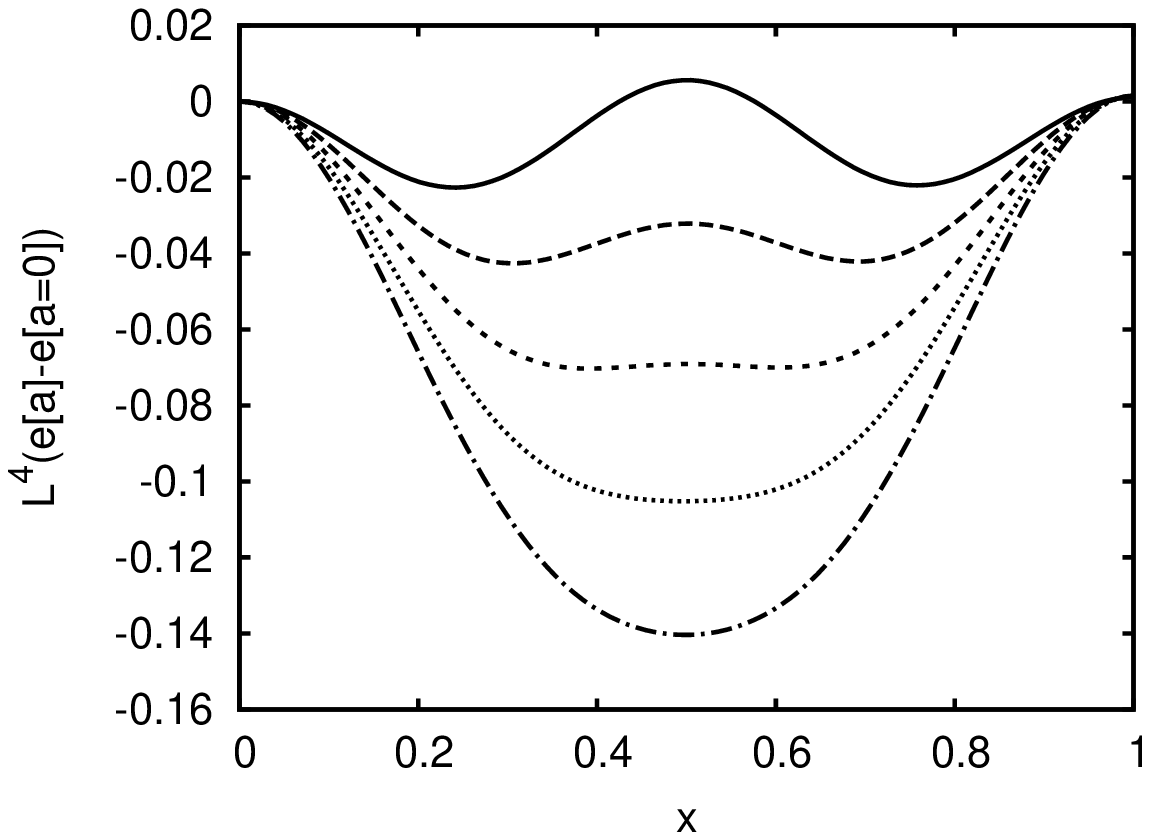}}
\subfigure[]{\label{fig7b}\includegraphics[width=0.85\linewidth]{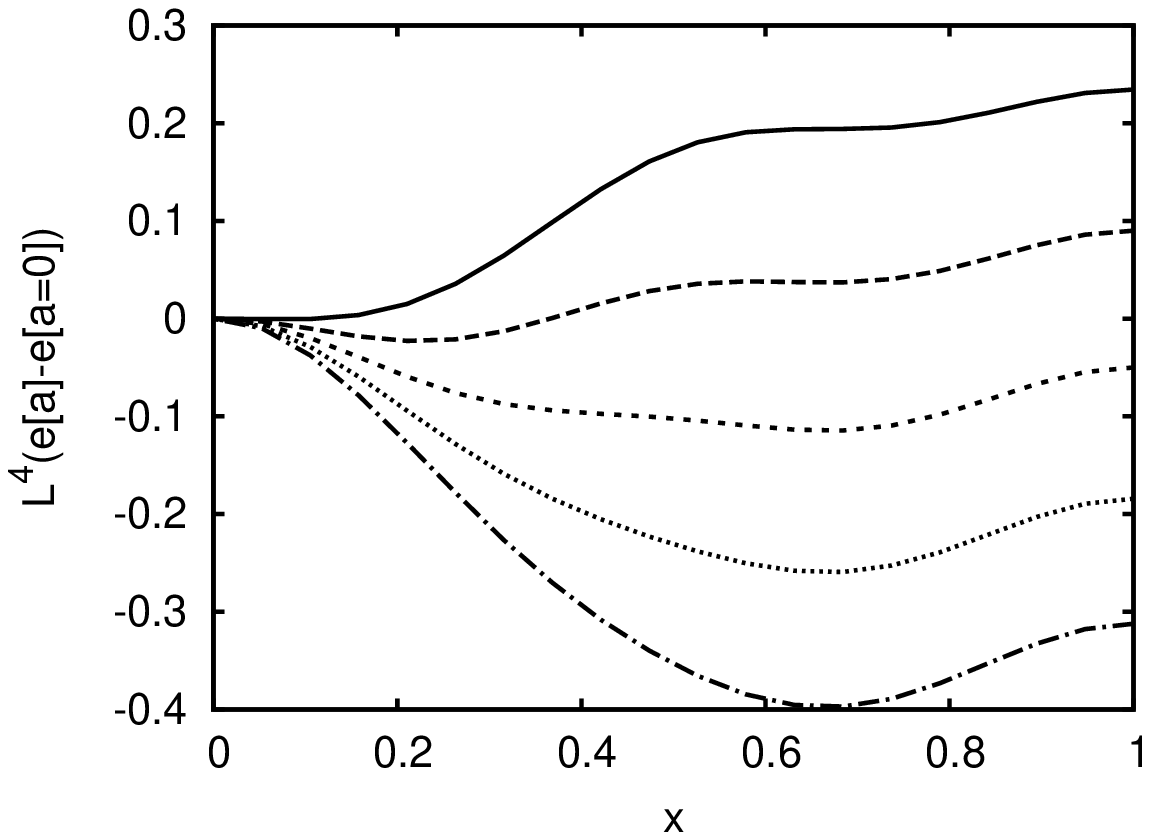}}
 \caption{The effective potential  (\ref{1875-ab}) for gauge group SU($2$) (a) and the $y=0$ ($a_8 = 0$) slice for SU($3$) (b) as a function of $x$ for different $L^{-1}$ in the range from $260$ to $290$ MeV in figure (a) and from  $260$ MeV to $300$ MeV in figure (b) (from bottom to top).  We extract the following phase transition temperatures: $T_c \approx 269$ MeV (SU($2$)) and $283$ MeV (SU($3$)).}
\label{fig5}
\end{figure}%

Alternatively we use for $\chi (p)$ the parameterization (\ref{1423-x1}) fitted to the numerical data and solve with this
$\chi (p)$ the gap equation (\ref{1360-x2}) for $\omega (p)$. To avoid UV divergencies in the numerical calculations we represent the potential $e (a, L)$ 
(\ref{1088-l6}) in the form
\begin{multline}
\label{1875-ab}
e (a, L) = \sum_\sigma \int \dbar[3] p \Big[ \omega (\vp_\sigma) - \omega_{\alpha = 1} (\vp_\sigma, \lambda = 0) \\
- u (\vp_\sigma) \chi_\text{IR} (\vp_\sigma) \Big] + e_{\alpha =1}(a,L,\lambda=0)
(a, L)\\
 - \sum_\sigma \int \dbar[3]\, p v (\vp_\sigma) \chi_\text{UV} (\vp_\sigma) \, .
\end{multline}
\begin{figure}
 \subfigure[]{\label{figN3-4a}\includegraphics[width=0.85\linewidth]{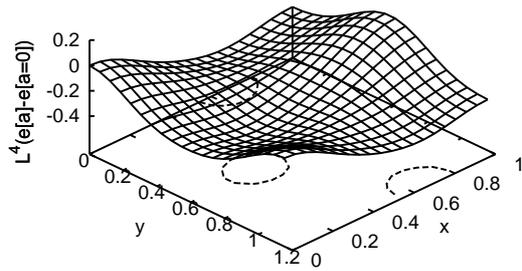}}
 \subfigure[]{\label{figN3-4b}\includegraphics[width=0.85\linewidth]{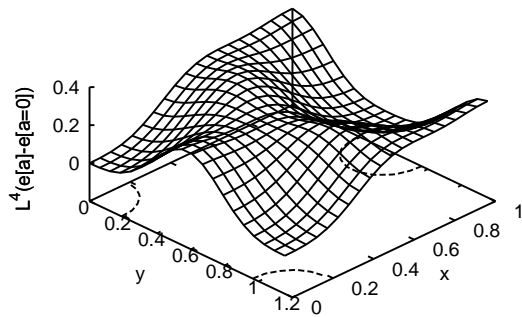}}
\caption{The effective potential (\ref{1803-aa}) for gauge group SU$(3)$ as a function of $x$ and $y$ (a) below and (b) above the phase transition temperature. In both plots the local minima are marked by a contour in the $xy$-plane.}
\label{figN3-4}
\end{figure}%
The first integral can be straightforwardly carried out numerically.
Using the explicit form of $\chi_\text{UV} (p)$ (\ref{1371-x4}) and $v (p)$ (\ref{1430-asa}) and the definition (\ref{1128-l10}) we may write the last term as
\begin{multline}
\label{1881-ab}
\sum_\sigma \int \dbar[3] p v (\vp_\sigma) \chi_\text{UV} (\vp_\sigma) \\
 =  \sum_\sigma \left[ \int \dbar[3] p 
\chi_\text{UV} (\vp_\sigma) - \lambda^n \int \dbar[3] p 
\frac{\chi_\text{UV} (\vp_\sigma)}{\lk \vp^2_\sigma + \lambda \rk^n} \right] \\
  = e^\chi_\text{UV} (a, L) - \lk \frac{\lambda}{M^2} \rk^n \int^\infty_0 \d t \, e_{\alpha = 1 - 2t - 2n} (a, L, \lambda) 
  \, ,
\end{multline}
where $e^\chi_\text{UV}$ is defined by Eqs.~(\ref{ymt-1527}) of appendix \ref{appB} and $e_\alpha  (a, L, \lambda)$ is given by Eq.~(\ref{1837-ii-13}).
The resulting effective potentials $e (a, L)$ (\ref{1875-ab}) are shown in Figs.~\ref{fig7a} and \ref{figN3-4} for the gauge group SU($2$) and SU($3$), respectively.
Fig.~\ref{fig7b} shows the $y=0$ cut through the SU($3$) potential shown in Fig.~\ref{figN3-4} for various temperatures $L^{-1}$. Figs.~\ref{fig7a} and \ref{fig7b} hardly differ from the potentials shown in Fig.~\ref{fig4} and \ref{figN3-3}, 
where the Gribov formula (\ref{1353-x1}) was assumed for $\omega (p)$ and the gap equation solved for $\chi (p)$ yielding 
(\ref{1365-x3}).
\begin{figure}
 \subfigure[]{\includegraphics[width=0.85\linewidth]{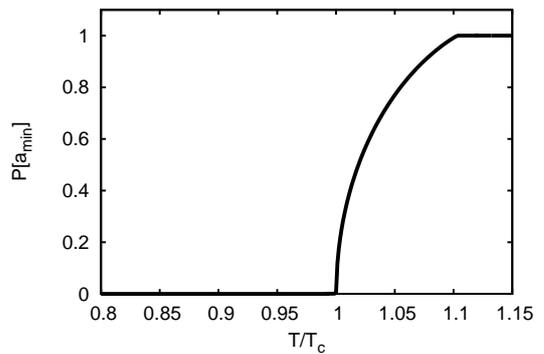}}
 \subfigure[]{\includegraphics[width=0.85\linewidth]{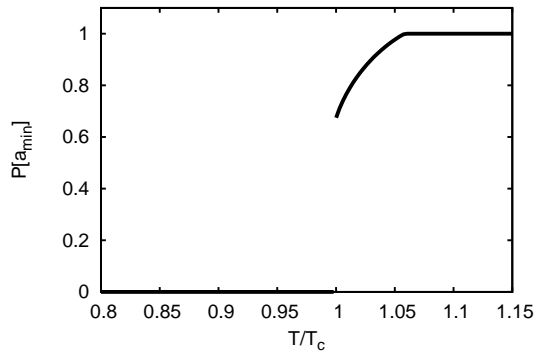}}
\caption{{The Polykav loop $\vev{P[a]}$  evaluated at the minimum $a = a_\text{min}$ of the full effective potential as a function of $T/T_c$ for the gauge groups SU($2$) (a) and SU($3$) (b).}}
\label{figN5}
\end{figure}
The critical temperature found from Fig.~\ref{fig5} is $T_c \approx 269$ MeV for SU($2$) and $T_c \approx 283$ MeV for SU($3$). These critical 
temperatures only slightly differ from the values found in Figs.~\ref{fig4both} from Eq.~(\ref{1803-aa}).
Within the given accuracy these values agree with the result of Ref.~\cite{R16} where the
variational approach to Yang--Mills theory in Coulomb gauge 
was directly extended to finite temperatures, studying the grand canonical ensemble. 
The present description has, however, the 
advantage that we do not need an additional variational ansatz for the density matrix of the gluons. 

Finally, Fig.~\ref{figN5} (a) and (b)  shows the Polyakov loop $P[a_\text{min}]$ as functions  of the temperature obtained from the minima $a_\text{min}$ of the full effective
potential for the gauge group SU($2$) and SU($3$), respectively. The order of the phase transition (second order for SU($2$) and first order for SU($3$)) is manifest in this figures. 

\section{\label{VII}Conclusions}

In the present paper we have studied the deconfinement phase transition by investigating Yang--Mills theory for one
compactified spatial dimension in the presence of an external background field directed along the compactified
dimension and living in the Cartan subalgebra. In this formulation the inverse of the length $L$ of the compactified dimension represents the temperature. 
The vacuum energy density as a function of the constant background field serves as the effective potential of the
Polyakov loop, the order parameter of confinement. We have calculated this potential within a variational approach by 
minimizing the energy density for given (value of the) background field and given compactified length $L$, using
a Gaussian type ansatz for the vacuum wave functional and a background gauge fixing. This resulted in a set of 
coupled DSEs for the gluon and ghost propagators. We have shown that these equations can be reduced to the corresponding 
DSEs in Coulomb gauge at $L^{- 1} = 0$ and in the absence of the background field. However, the momenta along the compactified 
dimension have to be taken at the discrete Matsubara frequencies and are shifted by the background field.
Using the zero-temperature results for the gluon and ghost propagators as input, from the effective potential
of the Polyakov loop we have extracted a critical temperature for the deconfinement phase transition of
$T_c \approx 269$ MeV for SU($2$) and $T_c \approx 283$ MeV for SU($3$). Within the given accuracy this values are consistent with the result of Ref.~\cite{R16}, where the
variational approach to Yang--Mills theory in Coulomb gauge was extended to non-zero temperature by making a
quasi-particle ansatz for the gluon density matrix and minimizing the free energy
 with respect to the  
quasi-gluon energy and the Fock basis. The presently calculated effective potential has also correctly revealed the order of the
deconfinement phase transition for SU($2$) and SU($3$).

In the present approach the deconfinement phase transition is entirely determined by the zero-temperature
propagators, which are defined as vacuum expectation values. This shows that the finite-temperature
behavior of the theory, in particular the dynamics of the deconfinement phase transition, must be fully encoded
in the vacuum wave functional, at least within the present truncation scheme.

The results obtained in the present paper are encouraging to extend the present approach to full QCD 
\cite{Pak:2011wu} at finite temperature and baryon density.

\begin{acknowledgments}
One of the authors (H.R.) acknowledges useful discussions with J.M. Pawlowski. We also thank D. Campagnari, M. Quandt and P. Watson
for a critical reading of the manuscript and useful comments. This work was supported by DFG under
contract DFG-Re 856/6-3, DFG-Re 856/9-1 and by BMBF under contract 06TU7199.
\end{acknowledgments}

\appendix

\section{\label{appA}Equations of motion}

In this Appendix we show that the equations of motion resulting from the constrained variational approach 
(\ref{603-G9}), i.e. the gap equation and the ghost DSE, can be consistently solved for propagators of the form 
assumed in Eqs.~(\ref{aa-852-z1}), (\ref{aa-855-z2}). As in the body of the paper we assume a constant background field living in the Cartan algebra. We first check that the form (\ref{aa-852-z1}) is consistent 
with the symmetries of the gluon propagator. 
The symmetry relation (\ref{dd-849-x2}) reads in momentum space 
\be
\label{dd-870-x6}
\cD^{ab}_{kl}  (\vp) = \cD^{ba}_{lk} (- \vp) \, .
\ee
Using 
\be
\label{dd-881-x8}
t^{ab}_{kl} (\vp) = t^{ba}_{lk} (- \vp)\,,
\ee
Eq.~(\ref{dd-859-x4}) can be rewritten in momentum space as
\be
\label{dd-886-x9}
t^{aa'}_{kk'} (\vp) \cD^{a' b'}_{k' l'} (\vp) t^{b' b}_{l' l} (\vp) = 
\cD^{ab}_{kl} (\vp) \, .
\ee
Defining the gluon propagator in the spherical basis by Eqs.~(\ref{488-21}) and using (\ref{460-*18}), (\ref{477-19})
the symmetry relation (\ref{dd-870-x6}) becomes in the 
spherical basis
\be
\label{dd-924-x15}
\cD^{\mu \nu}_{kl} (\vp) = (-)^{\mu + \nu} \cD^{- \nu, - \mu}_{lk} (- \vp) \, ,
\ee
while using (\ref{498-23}) Eq.~(\ref{dd-886-x9}) reads
\be
\label{dd-929-x16}
t^\sigma_{kk'} (\vp)  \cD^{\sigma \tau}_{k' l'} (\vp) t^\tau_{l' l} (\vp) = 
\cD^{\sigma \tau}_{kl} (\vp) \, .
\ee
Using Eq.~(\ref{498-23}) 
one easily convinces oneself that the diagonal ansatz (\ref{aa-852-z1})
is compatible with Eqs.~(\ref{dd-924-x15}) and (\ref{dd-929-x16}).

Next we analyse the equations of motion, i.e. the ghost DSE (\ref{aa-504-x2}) and 
the gap equation (\ref{643-29}) together with the expressions (\ref{aa-514-x4}) for the ghost self-energy and (\ref{yy-886}) for the curvature. We start with the ghost DSE (\ref{aa-504-x2}).
\subsection{The ghost self-energy}
Rewriting the ghost self-energy (\ref{y-782}) in momentum space
\begin{widetext}
\be
\label{y-791}
\Sigma (\vp)\\
 = \int \dbar q\, t^{aa'}_{kk'} (- \vq) t^{bb'}_{ll'} (\vq) \cD^{b' a'}_{l' k'} (\vq) 
\hat{T}_a \hat{d}_{k'} (\vp + \vq) \hat{G} (\vp + \vq) \hat{T}_b \hat{d}_l (\vp)  \, .
\ee
\end{widetext}
and switching to the spherical basis defined by Eqs.~(\ref{488-21}) we find
\begin{multline}
\label{ee-1075}
\Sigma^{\sigma \tau} (\vp) = (-)^{\sigma' + \tau'} \hat{T}_{- \sigma' \mu \sigma}^* \hat{T}_{- \tau' \nu \tau} \\
 \int \dbar[]q t^\nu_{ll'} (\vq) \cD^{\nu \mu}_{l' k'} (\vq) t^\mu_{k' k} (\vq) 
 d^\sigma_k (\vp + \vq) G^{\sigma' \tau'} (\vp  + \vq) d^\tau_l (\vp) \, ,
\end{multline}
where 
we have also defined the structure constant $\hat{T}^{ac}_b = \epsilon^{abc}$ in the
spherical basis 
\bea
\label{505-24}
\hat{T}_{\rho \sigma \tau} & = & \epsilon^{abc} \langle a | \rho \rangle \langle b | \sigma \rangle \langle c | \tau \rangle \nonumber\\
& = & \epsilon^{abc} e^a_\rho e^b_\sigma e^c_\tau = \ve_\rho  \lk \ve_\sigma \times \ve_\tau \rk \, .
\eea
Using Eq.~(\ref{482-20}) we obtain
\be
\label{511-25}
\hat{T}_{\rho \sigma \tau} = - \i \epsilon_{\rho \sigma \tau} \, .
\ee
Here the totally anti-symmetric tensor $\epsilon_{\rho \sigma \tau}$
 is defined as usual, however, for the indices $\sigma = - 1, 0, 1$. 
Since the indices $\rho, \sigma, \tau, ...$ take the values $0, \pm 1$ the totally antisymmetric
tensor $\varepsilon_{\rho \sigma \tau}$ is non-zero only for
\be
\label{ee-831}
\rho + \sigma + \tau = 0 \, .
\ee
Assuming in Eq.~(\ref{ee-1075}) color diagonal propagators (\ref{aa-852-z1}), (\ref{aa-855-z2}) we obtain
\begin{multline}
\label{864-oo}
\Sigma^{\sigma \tau} (\vp) = - \hat{T}_{- \sigma' \mu \sigma} \hat{T}_{- \sigma' \mu \tau} \\
\times \int \dbar[]q  t^\mu_{lk} (\vq)
\cD^\mu (\vq) d^\sigma_k (\vp + \vq) G^{\sigma'} (\vp + \vq) d^\tau_l (\vp) \, ,
\end{multline}
where we have used
\be
\label{871-oo}
t^\mu_{ll'} (\vq) t^\mu_{l' k'} (\vq) t^\mu_{k' k} (\vq) = t^\mu_{lk} (\vq) \, .
\ee
Since (no summation over $\rho, \mu$),
\be
\label{876-oo-G16}
\hat{T}^{*}_{- \rho \mu \sigma} \hat{T}_{- \rho \mu \tau} = \delta_{\sigma \tau} \left| T_{- \rho \mu \sigma} \right|^2 \, .
\ee
the ghost self-energy becomes indeed color diagonal
\be
\label{881-93}
\Sigma^{\sigma \tau} (\vp) = \delta^{\sigma \tau} \Sigma^\sigma (\vp)
\, ,
\ee
where
\begin{multline}
\label{y28}
\Sigma^\sigma (\vp)  = - \left| \hat{T}_{- \tau \mu \sigma}\right|^2 \\
\times \int \dbar q\, d^\sigma_l (\vp) t^\mu_{lk}
(\vq) d^\tau_k (\vp + \vq) \cD^\mu (\vq) G^\tau (\vp + \vq) \, .
\end{multline}
With the explicit form of $d^\sigma (\vp)$ (\ref{aa-903-G14}) we have
\bea
\label{yy-885}
d^\tau_k (\vp + \vq) & = & d^\mu_k (\vq) + d^{\tau - \mu}_k (\vp) \, .
\eea
Using furthermore that
\be
\label{yy-891}
d^\mu_k (\vq) t^\mu_{kl} (\vq) = 0
\ee
we can rewrite the ghost self-energy (\ref{y28}) as
\begin{multline}
\label{yy-896}
\Sigma^\sigma (\vp)= - \left| T_{- \tau \mu \sigma} \right|^2 \\
  \times \int \dbar q\, d^\sigma_l (\vp)
t^\mu_{lk} (\vq) d^{\tau - \mu} (\vp) \cD^\mu (\vq) G^\tau (\vq + \vp) \, .
\end{multline}
The antisymmetric tensor $\hat{T}_{- \tau \mu \sigma}$ is non-vanishing only for $\mu + \sigma - \tau = 0$.
Due to this sum rule the summation over $\tau$ can be explicitly carried out yielding 
with
\be
\label{956-mm}
\left| \hat{T}_{- (\mu + \sigma) \mu \sigma} \right|^2 = 1 
\ee
the final expression
\begin{multline}
\label{1124-ff-y1}
\Sigma^\sigma (\vp)=\\
 -  \sum_\mu \int \dbar q\, d^\sigma_l (\vp) t^\mu_{lk} (\vq) d^\sigma_l (\vp) 
\cD^\mu (\vq) G^{\sigma + \mu} (\vp + \vq) \, .
\end{multline}
Note this quantity is positive definite since the $d^\sigma_k (\vp)$ (\ref{aa-903-G14}) are purely imaginary. 
With Eq.~(\ref{881-93}) the ghost DSE (\ref{aa-504-x2}) becomes in the spherical basis
\be
\label{930-x4}
G^{\sigma^{- 1}} (\vp) = G^{\sigma^{- 1}}_0 (\vp) - \Sigma^\sigma (\vp) \, ,
\ee
where
\be
\label{935-x3}
G^{\sigma^{- 1}}_0 (\vp) = - \vd^\sigma (\vp)  \vd^\sigma (\vp) = \lk \vp - \ve_3 \sigma a \rk^2
\ee

and $\Sigma^\sigma (\vp)$ is defined by Eq.~(\ref{1124-ff-y1}). This shows that the diagonal ansatz, Eqs.~(\ref{aa-852-z1}), (\ref{aa-855-z2}) for the propagators is indeed compatible with the ghost DSE. What remains to be
done is to show that this ansatz is also compatible with the gap equation (\ref{643-29}). For this purpose 
we investigate first the curvature $\chi (\vq)$, which enters the gap equation.
\subsection{The curvature}

Fourier transforming the curvature (\ref{yy-886}) we find
\begin{multline}
\label{yy-897}
\chi^{ab}_{kl} (\vp) =\frac{1}{2} t^{aa'}_{kk'} (\vp) t^{bb'}_{ll'} (- \vp) \\
\times \int \dbar[]q
\tr \lk \hat{T}_{a'} \hat{d}_{k'} (\vp + \vq) G (\vp + \vq) \hat{T}_{b'} \hat{d}_{l'} (\vq) G (\vq) \rk \, .
\end{multline}
Rewriting this expression in the spherical basis (\ref{488-21})
and assuming a color diagonal ghost propagator (\ref{aa-855-z2}) we obtain
\begin{multline}
\label{ff-1179}
\chi^{\mu \nu}_{kl} (\vp) = -  \hat{T}^*_{- \tau \mu \sigma} \hat{T}_{- \tau \nu \sigma} \\
 \times \int \dbar[]q\, t^\mu_{kk'} (\vp) d^\tau_{k'} (\vq + \vp) d^\sigma_{l'} (\vq) t^\nu_{l' l} (\vp) 
G^\sigma (\vq) G^\tau (\vq + \vp) \, .
\end{multline}
With (\ref{876-oo-G16})
we find that the curvature is also color diagonal in the spherical basis
\be
\label{1191-ff}
\chi^{\mu \nu}_{kl} (\vp) = \delta^{\mu \nu} \chi^\mu_{kl} (\vp) \, .
\ee
Using
(\ref{yy-885})
and the sum rule for the indices of $\hat{T}_{- \tau \mu \sigma}$
the summation over $\tau$ can be trivially carried out, yielding 
\begin{multline}
\label{1206-ff}
\chi^\mu_{kl} (\vp) =\\
 -  \int \dbar  p\, t^\mu_{kk'} (\vp) d^\sigma_{k'}
(\vq) d^\sigma_{l'} (\vq) t^\mu_{l' l} (\vp) G^\sigma (\vq) G^{\mu + \sigma} (\vq + \vp) \, .
\end{multline}
Consider now the Lorentz structure 
of $\chi^\mu_{kl} (\vp)$. By its definition as derivative with respect to the (transversal) gauge field this
quantity has to be transverse as well. Indeed from the explicit expression (\ref{1206-ff}) it is seen that it satisfies
the transversality condition
\be
\label{1234-ff}
t^\mu_{m k} (\vp) \chi^\mu_{kl} (\vp) t^\mu_{ln} (\vp) = \chi^\mu_{mn} (\vp) \, .
\ee
From this we can conclude that it has the form\footnote{The transversal projector $t^\mu_{kl} (\vp)$ will
explicitly show up after the angular integral in (\ref{1206-ff}) is carried out. Furthermore, the factor $g^2$ has been explicitly
included so that $\chi^\mu (q)$ reduces for $\mu = 0$ to the curvature defined in Ref.~\cite{Feuchter:2004mk}, see Sect.~\ref{VF}. }
\be
\label{1240-ff}
\chi^\mu_{kl} (\vp) = \frac{1}{g^2}  t^\mu_{kl} (\vp) \chi^\mu (\vp) \, ,
\ee
where the scalar curvature $\chi^\mu (\vp)$ can be obtained as
\be
\label{1245-ff}
\chi^\mu (\vp) = \frac{g^2}{d - 1} t^\mu_{lk} (\vp) \chi^\mu_{kl} (\vp) \, .
\ee
Here $d$ is the number of spatial dimensions and we have used
\be
\label{1192-ad}
\hat{t}^\mu_{kk} (\vp) = d - 1 \, .
\ee
Inserting here the explicit expression (\ref{1206-ff}) and using
$
\label{1250-ff}
t^\mu_{l' l} (\vp) = t^\mu_{ll'} (\vp)$
we obtain
\begin{multline}
\label{1266-ff-x1}
\chi^\mu (\vp) = - \frac{g^2}{2 (d - 1)}\\
 \times \sum_\sigma \int \dbar[]q d^\sigma_k (\vq) t^\mu_{kl} (\vp) d^\sigma_l (\vq)
G^\sigma (\vq) G^{\mu + \sigma} (\vp + \vq) \, .
\end{multline}
Eqs.~ (\ref{1191-ff}) and (\ref{1240-ff}) show that in the spherical basis the curvature $\chi (q)$ is not only color diagonal
but has also the same Lorentz structure as assumed for the gluon propagator (\ref{aa-852-z1}).


\subsection{The gap equation}

Consider now the gap equation (\ref{643-29}), which after Fourier transformation becomes
\begin{multline}
\label{aa-836}
\frac{g^2}{4} \cD^{- 1} (\vp) \cD^{- 1} (\vp)\\
 = \frac{1}{g^2} \lk - \hat{t} (\vp) 
\hat{\vd} (\vp) \hat{\vd} (\vp) \hat{t} (\vp) \rk + g^2
\chi (\vp) \chi (\vp) \, .
\end{multline}
This is still a matrix equation both in the adjoint color and Lorentz indices. 
Rewriting this equation in the spherical basis defined by Eq.(\ref{488-21}) and using Eqs.~(\ref{493-22}) and (\ref{1191-ff}) 
we observe that the gluon propagator $\cD (\vp)$ has indeed
to be color diagonal and obtain
\begin{multline}
\label{1015-y1}
\frac{g^4}{4} \cD^\sigma (\vp)^{- 1} \cD^\sigma (\vp)^{- 1}\\
= \lk  - t^\sigma (\vp) 
\vd^\sigma (\vp) \vd^\sigma (\vp) t^\sigma (\vp) \rk +
g^4 \chi^\sigma (\vp) \chi^\sigma (\vp) \, .
\end{multline}
Here $\cD^{\sigma} (\vp)^{- 1}$ and $\chi^\sigma (\vp)$, like $t^\sigma (\vp)$,
 are still Lorentz matrices. This equation is compatible with the 
Lorentz structure of the gluon propagator assumed in (\ref{aa-852-z1}), which resulted in   the Lorenz structure
(\ref{1240-ff}) of the curvature. Thus the ans\"atze (\ref{aa-852-z1}), (\ref{aa-855-z2}) for the propagators
are also
compatible with the gap equation. We have thus shown that the equation of motions resulting from the constraint variational
principle $\vev { H }_a \to \min, \, \vev{ A }_a = a$ in the background gauge (\ref{ex-7}) allow for color diagonal
solutions. For such solutions we find for the energy density (\ref{650-30}) at the stationary point
\begin{multline}
\label{1026-pp}
\vev {H_K + H^A_B }_a \\
= g^2 V \sum_\sigma \int \dbar[]q \lk \frac{1}{2} \cD^\sigma (\vq)^{- 1}
- \chi^\sigma (\vq) \rk \, ,
\end{multline}
where $V$  is the spatial volume and 
 $\cD^\sigma (p)$ and $\chi^\sigma (p)$ are still Lorentz matrices. Carrying out the trace over the Lorentz indices thereby
using Eqs.~(\ref{aa-852-z1}) and (\ref{1240-ff}) we obtain
\begin{multline}
\label{1241-qq}
\vev { H_K + H^A_B }_a \\
= (d - 1) V \sum_\sigma \int \,  \dbar[]q \lk \frac{g^2}{2} \cD^\sigma (\vq)^{- 1} - \chi^\sigma (\vq) \rk \, ,
\end{multline}
where $\cD^\sigma (\vp)$ and $\chi^\sigma (\vp)$ are now Lorentz scalars. (Note by Eq.~(\ref{1240-ff}) the Lorentz tensor $\chi^\sigma_{kl} (\vp)$ and the Lorentz
scalar $\chi^\sigma (\vp)$ differ by a factor of $g^2$.)


\section{\label{appB}The energy density}
Below we calculate the contribution of the generic gluon energy (\ref{1128-l10}) to the energy density (\ref{1118-l9}). Representing 
\be
\label{1134--l11}
\lk \sqrt{p_\sigma^2 + \lambda} \rk^\alpha = \lk \vp^2_\perp + \lk p_n - \vec{\sigma} \va \rk^2 + \lambda \rk^{\alpha /2}
\ee
by a proper-time integral\footnote{We assume here that $\alpha$ is such that the proper-time integral exists, otherwise 
$\int^\infty_0 \d \tau$ has to be replaced by $\int^\infty_{1/\Lambda^2} \d \tau$, where $\Lambda$ is a 
UV cut-off.}
\be
\label{1141-l12}
(p_\sigma^2 + \lambda)^{\alpha/2} = \frac{1}{\Gamma \lk - \frac{\alpha}{2} \rk} \int^\infty_0 d \tau \tau^{- 1 - \frac{\alpha}{2}} 
\e^{- \tau \lk p_\sigma^2 + \lambda \rk} \, 
\ee
we find for the energy density (\ref{1118-l9}) 
\begin{multline}
\label{1772-ii-4-13}
e_\alpha (a, L, \lambda) = \frac{M^{1 - \alpha}}{4 \pi} \frac{1}{\Gamma \lk - \frac{\alpha}{2} \rk} \\
\times \sum_\sigma \frac{1}{L} \sum_n  
\int^\infty_0 \d x \int^\infty_0 \d \tau \tau^{- 1 - \frac{\alpha}{2}} \e^{- \tau \left[ x + \lk k_n - \vec{\sigma} \va \rk^2 \right] } 
\e^{- \lambda \tau}  .
\end{multline}
Here we have carried out the integral over the azimutal angle of $\vp_\perp$ and set $x = \vp^2_\perp$. 
The integral over $x$ can be carried out yielding  a factor $\tau^{- 1}$. By means of the Poisson 
resummation formula one derives the relation
\be
\label{1784-ii-6-14}
\frac{1}{L} \sum_n f (p_n) =  \frac{1}{2 \pi} \int^\infty_{- \infty} \d z f (z) \sum_n \e^{\i n z L} 
\, , \quad  p_n = \frac{2 n \pi}{L} \,.
\ee

Using this relation in Eq.~(\ref{1772-ii-4-13}) and performing the integration over $z$ we obtain
\begin{multline}
\label{1795-ii-8-15}
e_\alpha (a, L, \lambda) = \frac{M^{1 - \alpha}}{(4 \pi)^{3/2}} \frac{1}{\Gamma \lk - \frac{\alpha}{2} \rk}\\
\times \sum_\sigma \sum^\infty_{n = - \infty}
\e^{\i n \vec{\sigma}  \va L} \int^\infty_0 \d \tau \tau^{- \frac{5}{2} - \frac{\alpha}{2}} \e^{- \frac{1}{\tau} \lk 
\frac{n L}{2} \rk^2} \e^{- \lambda \tau} \, .
\end{multline}
Obviously the proper-time integral is not well-defined for $\lambda = 0$ and
 $n = 0$. Fortunately, this term drops out by subtracting 
the energy density at vanishing background field
\begin{multline}
\label{1802-ii-9}
\bar{e}_\alpha (a, L, \lambda) \coloneq e_\alpha (a, L, \lambda) - e_\alpha (a=0, L, \lambda)\\
 = \, \frac{M^{1 - \alpha}}{(4 \pi)^{3/2}} \frac{2}{\Gamma \lk - \frac{\alpha}{2} \rk}\\
\times  \sum_\sigma  \sum^\infty_{n = 1}  \left[ \cos (n \vec{\sigma} \va L) - 1 \right] \int^\infty_0 \d \tau \tau^{- \frac{5}{2} - \frac{\alpha}{2}} 
\e^{- \frac{1}{\tau} \lk \frac{n L}{2} \rk^2 - \lambda \tau}.
\end{multline}
With a change of variable $s = \lk n L / 2 \rk^2 / \tau$ in
the proper-time integral 
\begin{multline}
\label{1821-ii-10}
\int^\infty_0 d \tau \tau^{- \frac{5}{2} - \frac{\alpha}{2}} \e^{- \frac{1}{\tau} \lk \frac{n L}{2} \rk^2} \e^{- \lambda \tau}\\
= \lk \frac{2}{n L} \rk^{3 +  \alpha} \int^\infty_0 \d s \,s^{\frac{1}{2} + \frac{\alpha}{2}} \e^{- \frac{\lambda}{s} \lk \frac{n L}{2} \rk^2 } \, 
\end{multline}
we find for the energy density (\ref{1802-ii-9}) after some trivial algebraic manipulations
\begin{widetext}
\be
\label{1486-qq}
\bar{e}_\alpha (a, L, \lambda)  = - 4 (4 \pi)^{- 3/2} \frac{M^{1 - \alpha}}{\Gamma \lk - \frac{\alpha}{2} \rk} 
 \sum_\sigma \sum^\infty_{n = 1} \lk \frac{2}{n L} \rk^{\alpha + 3} \sin^2 \lk  n \frac{\vec{\sigma} \va  L}{2} \rk 
\int^\infty_0 \d s s^{\frac{1}{2} + \frac{\alpha}{2}} \e^{- s} \e^{- \frac{\lambda}{s} \lk \frac{n L}{2} \rk^2} \, .
\ee
\end{widetext}
For $\lambda = 0$ the remaining integral can be expressed by Euler's Gamma-function
\begin{multline}
\label{1830-ii-11}
\bar{e}_{\alpha} (a, L, \lambda = 0)  = -  \frac{4 M^{1 - \alpha}}{(4 \pi)^{3/2}} \frac{\Gamma \lk \frac{3}{2} + \frac{\alpha}{2} \rk}{\Gamma \lk
- \frac{\alpha}{2} \rk} \\
\times \sum_\sigma  
\lk \frac{2}{L} \rk^{3 + \alpha} \sum^\infty_{n = 1} \frac{1}{n^{3 + \alpha}} \sin^2 \lk n \vec{\sigma} {\va L}/{2} \rk \, .
\end{multline}
Obviously, the vanishing roots $\vec{\sigma} = (0, 0)$ do not contribute to the potential difference $e_\alpha (a) - e_\alpha (a = 0)$.
Furthermore, the negative roots give the same contribution as the positive ones. We can therefore restrict the summation over $\sigma$
to the positive roots $\sigma > 0$ and take care of the contribution of the negative roots by an additional factor $2$. This yields
\begin{multline}
\label{1837-ii-12-alt}
\bar{e}_{\alpha} (a, L, \lambda = 0)   = - 8 \frac{M^{1 - \alpha}}{(4 \pi)^{3/2}} \frac{\Gamma \lk 
\frac{3}{2} + \frac{\alpha}{2} \rk}{\Gamma \lk
- \frac{\alpha}{2} \rk}  \lk \frac{2}{L} \rk^{3 + \alpha} \\
\times \sum_{\vec{\sigma} > 0} \sum^\infty_{n = 1} \frac{1}{n^{3 + \alpha}} \sin^2 \lk n
 \vec{\sigma} \va L /2 \rk \, .
\end{multline}
For $\lambda \neq 0$ the remaining proper-time integral in Eq.~(\ref{1486-qq}) cannot be taken analytically but can
be expressed by a modified Bessel function $K_\nu(z)$ 
\begin{multline}
\label{1666-ac}
\int_0^\infty \d s s^{-1 - t}\e^{-s} \exp \left[ -\frac{\lambda}{s} \left(\frac{n L}{2} \right)^{2 }  \right]\\
= 2 \left(\frac{2}{L n \sqrt{\lambda}}\right)^{t} K_{ t}(n L \sqrt{\lambda})\, ,
\end{multline}
yielding 
\begin{multline}
\label{1837-ii-13-alt}
\bar{e}_{\alpha} (a, L, \lambda )  = - 8 \frac{M^{1 - \alpha}}{(4 \pi)^{3/2} \Gamma \lk
- \frac{\alpha}{2} \rk} \\
\times \sum_\sigma \sum^\infty_{n = 1}  \lk \frac{2 \sqrt{\lambda}}{n L} \rk^{\frac{\alpha}{2}+\frac{3}{2}} \sin^2 
\left( \vec{\sigma} \va L /2 \right) K_{-\frac{\alpha}{2}-\frac{3}{2}}(n L \sqrt{\lambda})\, .
\end{multline}
As in the previous case $\lambda =0$, the summation over $\sigma$ can be restricted to $\sigma > 0$ by including a factor of 2.

To eliminate the UV divergencies from the energy density $e (a, L)$ (\ref{1088-l6}) 
we also need to isolate the contribution from the UV behavior of $\chi $. The UV form of $\chi$ is given by Eq.~(\ref{1371-x4}). Using
\be
\label{1900-x2}
\frac{1}{\ln \frac{p_\sigma^2 + \lambda}{M^2}} = \int^\infty_0 \d t \lk \frac{M^2}{p_\sigma^2  + \lambda} 
\rk^t , \quad p_\sigma^2 + \lambda > M^2  \, 
\ee
and the definition (\ref{1128-l10}) of $\omega_\alpha (p, M, \lambda)$ we have
\be
\label{ymt-1527}
\chi_\text{UV} (p) = \int^\infty_0 \d t \,\omega_{\alpha = 1 - 2t} (p, \lambda) \, .
\ee
With this relation the (negative of the) energy contribution generated by the UV-behavior of $\chi$ reads
\begin{align}
\label{1293-qq-z1}
e^\chi_\text{UV} (a, L) \coloneq& \sum_\sigma \int \dbar p \,\chi_\text{UV} (p_\sigma)\\
=& \int^\infty_0 \d t \,e_{\alpha = 1 - 2t} (a, L, \lambda) \nonumber \, ,
\end{align}
\begin{widetext}
where $e_\alpha (a, L, \lambda)$ is defined in Eq.~(\ref{1772-ii-4-13}).
Subtracting the $a = 0$ contribution and using Eq.~(\ref{1837-ii-13}) we obtain 
\be
\label{1664-x}
e^\chi_\text{UV} (a, L) - e^\chi_\text{UV} (a = 0, L)  = - \frac{8}{(4 \pi)^{3/2}} \int^\infty_0 \d t \, \frac{M^{2 t}}{\Gamma \lk t - \frac{1}{2} \rk} \lk\frac{\sqrt{\lambda}}{L} \rk^{2- t}  
 \sum_\sigma \sum^\infty_{n = 1} \lk \frac{2}{n} \rk^{2 - t}   {\sin^2} \lk n \vec{\sigma} \va \frac{L}{2} \rk K_{t - 2} \lk n L \sqrt{\lambda} \rk
\ee
and 
after restricting the summation over $\sigma$ to positive roots
\be
\label{1928-ll}
e^\chi_\text{UV} (a, L) - e^\chi_\text{UV} (a = 0, L)  = - \frac{16}{(4 \pi)^{3/2}} \int^\infty_0 \d t  \,\frac{M^{2 t}}{\Gamma \lk t - \frac{1}{2} \rk}\lk\frac{\sqrt{\lambda}}{L} \rk^{2- t} 
 \sum_{\sigma > 0} \sum^\infty_{n = 1} \lk \frac{2}{n} \rk^{2 - t}   {\sin^2} \lk n \vec{\sigma} \va \frac{L}{2} \rk K_{t - 2} \lk n L \sqrt{\lambda} \rk
 \, .
\ee
\end{widetext}


\end{document}